\documentclass[11pt,a4paper,aps,tightenlines,
preprintnumbers,nofootinbib,showkeys,superscriptaddress]{revtex4-1}
\pdfoutput=1
\usepackage{fullpage}
\usepackage{float}
\usepackage[normalem]{ulem}
\usepackage{amsfonts}
\usepackage{amsmath}
\usepackage{slashed}
\usepackage{mathrsfs}
\usepackage[scr=rsfs,cal=boondox]{mathalfa}
\usepackage{amssymb}
\usepackage{graphicx}
\usepackage{makeidx}
\usepackage{eepic}
\usepackage{dsfont}
\usepackage{epsfig}
\usepackage{latexsym}
\usepackage{mathtools}
\usepackage[dvipsnames]{xcolor}
\usepackage{float}
\usepackage{multirow}
\usepackage[export]{adjustbox}
\usepackage{xurl,hyperref}
\usepackage{enumitem}
\hypersetup{colorlinks=true,citecolor=red,linkcolor=NavyBlue,urlcolor=NavyBlue}
\usepackage[utf8]{inputenc}
\usepackage[caption=false]{subfig}
 
\usepackage{natbib}
\usepackage{mathptmx}
\newcommand{\M}{\mathcal{M}}
\def\ba{\begin{array}}       
	\def\ea{\end{array}}

\def\beq{\begin{eqnarray}}
\def\eeq{\end{eqnarray}}
\begin{document}

	\title{Electroweak renormalization of neutralino-Higgs interactions at one-loop and its impacts on spin-independent direct detection of Wino-like dark matter}
	\author{Subhadip Bisal}
	\email{subhadip.b@iopb.res.in}
	\affiliation{ Institute of Physics, Sachivalaya Marg, Bhubaneswar, 751 005, India}
	
	\affiliation{Homi Bhabha National Institute, Training School Complex, Anushakti Nagar, Mumbai 400 094, India}
	
	\author{Arindam Chatterjee}
	\email{arindam.chatterjee@snu.edu.in}
	\affiliation{Shiv Nadar IoE Deemed to be University, Gautam Buddha Nagar, Uttar Pradesh, 201314, India}
	
	\author{Debottam Das}
	\email{debottam@iopb.res.in}
	\affiliation{ Institute of Physics, Sachivalaya Marg, Bhubaneswar, 751 005, India}
	\affiliation{Homi Bhabha National Institute, Training School Complex, Anushakti Nagar, Mumbai 400 094, India}
	
        \author{Syed Adil Pasha}
	\email{sp855@snu.edu.in}
	\affiliation{Shiv Nadar IoE Deemed to be University, Gautam Buddha Nagar, Uttar Pradesh, 201314, India}

	\date{\today}

\begin{abstract}
    \noindent
A Wino-like neutralino dark matter (DM) in the form of the lightest supersymmetric particle (LSP) has been considered one of the popular paradigms that can naturally accommodate {\it new physics} at a relatively higher scale, typically beyond the reach of the LHC. The constraint on the DM relic density typically implies a lightest neutralino mass $\simeq 2$ TeV. Its observational signature through nuclear recoil experiments, specifically involving DM-nucleon spin-independent (SI) scattering, is not impressive, following its high masses and tiny Higgsino fractions. 
The theoretical calculations can be improved when we compute all the one-loop electroweak (EW) corrections to the three-point vertices for the neutralino (Wino)-Higgs interactions, 
which in turn boosts the DM-nucleon scattering cross-sections through the SM-like Higgs exchange.
Importantly, we include the counterterm contributions.
In addition, we incorporate the other next-to-leading order (NLO) EW DM-quark and DM-gluon interactions present in the literature to calculate the DM-nucleon scattering cross-sections. 
With the improved and precise theoretical estimates, DM-nucleon scattering cross-sections may increase or decrease significantly by more than $100\%$ compared to leading order (LO) cross-sections in different parts of the parameter space. 
\end{abstract}
 
	\maketitle
	\section{INTRODUCTION}

The minimal supersymmetric standard model (MSSM)
is the most popular extension of the standard model (SM), though, in the LHC era, the situation has become distressing following the null searches for squarks and gluino. 
The masses of squarks and gluino are typically assumed at $\geq$ 2.5~TeV to cope with the LHC constraints~\cite{ATLAS:2020syg, CMS:2019zmd}. 
A lighter neutralino or chargino (collectively referred to as electroweakino) still stands as the best pledge for the TeV scale supersymmetry (SUSY).  
With $R$-parity conservation, the weakly interacting lightest neutralino, $\tilde \chi_1^0$, which turns out to be the lightest supersymmetric particle (LSP) in the MSSM parameter space, serves as the cold dark matter (CDM) in the Universe.
The {\it lightness} of the electroweak (EW) states may boost the matrix elements corresponding to the scattering of the dark matter (DM) on nuclei. 
For instance, for a predominantly Higgsino-like LSP, the next-to-leading order (NLO) corrections to the spin-independent (SI) direct detection (DD) (SI-DD) cross-section can be as large as $100\%$~\cite{Bisal:2023fgb}. Similarly, for Higgsino- and Wino-like LSPs, NLO corrections to the spin-dependent (SD) DD cross-sections may exceed $100\%$~\cite{Bisal:2024ezn}. In contrast, for a Bino-dominated LSP with a minimal Higgsino component, the SI-DD cross-section is enhanced by approximately $20-30\%$ relative to the leading order (LO) value~\cite{Bisal:2023iip}. In the same parameter space, the indirect search prospects of sub-TeV Bino-like DM is studied in Ref.~\cite{Chattopadhyay:2024qgs}.

In the context of Wino-like DM (a natural candidate in high-scale SUSY models like anomaly-mediated supersymmetry breaking (AMSB)~\cite{Randall:1998uk,Giudice:1998xp}, or
can be realized in string-inspired 
scenarios~\cite{Brignole:1993dj,Casas:1996wj}), 
 $\tilde \chi_1^0$ possesses an isospin of $\frac{1}{2}$, enabling it to undergo pair annihilation into either pair of $W$ bosons or fermions via its gauge couplings to the $W$ boson. 
The resulting DM relic density is controlled mainly by the mass of the LSP, and for Wino-like ($\widetilde{W}$) $\tilde{\chi}_1^0$, it corresponds to a Wino mass of $1.9-2.3$ TeV~\cite{Chattopadhyay:2006xb, Chakraborti:2017dpu} (Sommerfeld enhancement (SE) can change the Wino mass to $2.7-3$ TeV~\cite{Hisano:2006nn, Mohanty:2010es}),
which can comply with the observed relic abundance data~\cite{WMAP:2012nax, Planck:2018vyg}~\footnote{In case $\tilde{\chi}_1^0$ also accommodates the Higgsino or the Bino component, the observed relic abundance prefers a relatively lighter LSP~\cite{Chattopadhyay:2009fr}.}\,:
 \begin{equation}
   \Omega_{\rm DM}h^2=0.1198\pm 0.0012.
   \label{eq:relic}
 \end{equation}

The SI searches for Wino-like DM critically depend on the Higgsino component that may arise when diagonalizing the neutralino mass matrix. Since the Higgs coupling to the LSP pair is proportional to the product of their Higgsino and gaugino components, one finds a vanishingly small $\tilde{\chi}_1^0$-nucleon scattering cross-section at the LO when the Higgsino fraction is negligible. 
Thus, the
$\tilde{\chi}_1^0$-nucleon scattering at NLO (involving the SM and beyond SM or BSM particles) with quarks and gluons has been of interest in the past
Ref.~\cite{Drees:1992rr, Drees:1993bu, Hisano:2004pv, Hisano:2010ct, Hisano:2010fy, Hisano:2012wm, Hisano:2011cs, Hisano:2015rsa, Klasen:2016qyz, Ellis:2023ndh} (for a general EW WIMP, see Ref.~\cite{Essig:2007az, Cirelli:2005uq}). In regard to EW corrections,
the authors mainly considered~\cite{Hisano:2004pv, Hisano:2010fy, Hisano:2012wm, Ellis:2023ndh, Hisano:2011cs, Hisano:2015rsa}:
(i) the three-point $\tilde{\chi}_1^0$-$\tilde{\chi}_1^0$-Higgs vertex corrections mediated by $\tilde \chi_1^\pm$ and $W^\pm$, (ii) NLO corrections to $\tilde{\chi}_1^0\tilde{\chi}_1^0 q \bar{q}$ and
$\tilde{\chi}_1^0\tilde{\chi}_1^0 g g$ vertices,  which, in particular involve quark and gluon \textbf{twist-2} operators 
(see Sec.~\ref{subsec:presentstatus} for details).

In this work, we go one step further and incorporate all the one-loop triangular diagrams involving EW particles that may contribute to the LSP-nucleon SI interactions through $\tilde{\chi}_1^0$-$\tilde{\chi}_1^0$-Higgs vertex for a Wino-like $\tilde{\chi}_1^0$~\footnote{Several studies have suggested that indirect detection experiments, particularly H.E.S.S. and Fermi-LAT, constrain both thermal Winos and non-thermal Wino DM~\cite{Cohen:2013ama, Fan:2013faa, Hryczuk:2014hpa, Bhattacherjee:2014dya, Baumgart:2014saa, Beneke:2016jpw, Baumgart:2018yed, Ando:2019rgx, Rinchiuso:2020skh}. The stringency of the constraints is subject to the uncertainties in the DM density profiles.}. 
Assuming sfermions to reside at the multi-TeV scale, the dominant contributions to the SI-DD cross-section primarily arise from Higgs exchange processes.
As $\tilde{\chi}_1^0$ can accommodate some Higgsino components, the tree-level $\tilde{\chi}_1^0$-$\tilde{\chi}_1^0$-Higgs vertex appears, and consequently, the renormalization of this vertex becomes necessary to have a meaningful result.
We present the necessary details of the renormalization of the neutralino-Higgs vertices and include the contributions from the counterterms for the accurate estimation of the SI-DD cross-sections, which, to the best of our knowledge, is not present in the literature. 

The manuscript is organized as follows. In Sec.~\ref{sec:effectivelagrng}, we briefly discuss the neutralino-nucleon scattering, 
while the relevant parts of the $\tilde\chi_1^0$-$\tilde\chi_1^0$-Higgs vertices have been summarized.
Sec.~\ref{sec:presentstatus} presents an improved analysis of the SI-DD of Wino-like DM. In particular, Sec.~\ref{subsec:presentstatus} discusses the present status related to the theoretical developments and our objective of the work; Sec.~\ref{subsec:ver} and Sec.~\ref{subsec:renormalization} present the one-loop triangular topologies for $\tilde{\chi}_1^0$-$\tilde{\chi}_1^0$-Higgs vertex corrections and the details of the renormalization. 
In Sec.~\ref{sec:improvementMicro}, we discuss the chronology of calculating the neutralino-nucleon scattering matrix elements and their implementation in $\mathtt{MicroMEGAs}$. 
Additionally, in this section, after validating our code with the existing literature, we highlight the salient observations.
We present the numerical results in Sec.~\ref{sec:numericalres} and finally conclude in Sec.~\ref{sec:conclusion}.

\section{A Brief Reprisal of $\tilde{\chi}_1^0 (\widetilde{W})-$Nucleon Scattering}
\label{sec:effectivelagrng}
\textbf{$\bullet$ The neutralino-Higgs interactions\,:} The Lagrangian for the neutralino-neutralino-scalar interaction in the MSSM can be expressed as~\cite{Drees:2004jm},
\begin{align}
\mathcal{L}_{\tilde{\chi}_\ell^0\tilde{\chi}_n^0\phi} \supset& \,\,\frac{g_2}{2}h \bar{\tilde{\chi}}_n^0 \Big[\mathbf{P_L}\Big(Q_{\ell n}^{\prime\prime *}s_\alpha + S_{\ell n}^{\prime\prime *}c_\alpha\Big) + \mathbf{P_R}\Big(Q_{n\ell}^{\prime\prime}s_\alpha + S_{n\ell}^{\prime\prime}c_\alpha\Big)\Big]\tilde{\chi}_{\ell}^0 \nonumber\\
-& \frac{g_2}{2}H \bar{\tilde{\chi}}_n^0 \Big[\mathbf{P_L}\Big(Q_{\ell n}^{\prime\prime *}c_\alpha - S_{\ell n}^{\prime\prime *}s_\alpha\Big) + \mathbf{P_R}\Big(Q_{n\ell}^{\prime\prime}c_\alpha - S_{n\ell}^{\prime\prime}s_\alpha\Big)\Big]\tilde{\chi}_{\ell}^0\nonumber\\
-&i\frac{g_2}{2}A\bar{\tilde{\chi}}_n^0 \Big[\mathbf{P_L} \Big(S^{\prime\prime *}_{\ell n}c_\beta -Q^{\prime\prime *}_{\ell n}s_\beta\Big) + \mathbf{P_R} \Big(Q^{\prime\prime}_{n\ell}s_\beta -S^{\prime\prime}_{n\ell}c_\beta\Big)\Big] \tilde{\chi}_{\ell}^0~.
\label{Eq:lochichihi}
\end{align} 

Here, $c_\alpha=\cos\alpha$, $s_\alpha=\sin\alpha$, $c_\beta=\cos\beta$, $s_\beta=\sin\beta$, $\tilde{\chi}_{\ell,n}^0$'s \big($\ell, n=1,...,4$\big) are the neutralinos, $\phi=h_i,A$, with $h_i$ refers to an SM-like scalar $h$ or a heavy CP-even Higgs boson $H$, respectively. The neutral CP-odd Higgs is denoted by $A$, $\alpha$ represents the Higgs mixing angle, $\beta$ is the inverse tangent of the ratio of the vacuum expectation values (vevs) of the CP-even neutral Higgs bosons, and $g_2$ is the $SU(2)_L$ gauge coupling strength. Similarly, $\mathbf{P_{L,R}}= \frac{1\mp \gamma 5}{2}$ as usual. The expressions for
$Q^{\prime\prime}_{n\ell}$ and $S^{\prime\prime}_{n\ell}$ are given in Ref.~\cite{Drees:2004jm}. 

In the parameter space, where $\tilde \chi_1^0$ becomes Wino-like ($M_2<<|\mu|, M_1$) or Higgsino-like ($|\mu|<<M_1, M_2$), the mass eigenvalues for the lightest neutralino and the lightest chargino are close to each other. For Wino-like neutralino, one can write~\cite{Hisano:2004pv},

\begin{align}
m_{\tilde{\chi}_1^0} = M_2 + \frac{M_W^2}{M_2^2-\mu^2} \big(M_2 + \mu s_{2\beta}\big) + ...~,\\
m_{\tilde{\chi}_1^\pm} = M_2 + \frac{M_W^2}{M_2^2-\mu^2} \big(M_2 + \mu s_{2\beta}\big) + ...~.
\end{align}

It is important to note that the masses mentioned in the above expressions are tree-level masses. Loop corrections, which can be substantial, have been systematically computed at the one-loop level in Ref.~\cite{Aoki:1982ed, Pierce:1993gj, Pierce:1994ew, Pierce:1996zz}, and partial two-loop results can be found in Ref.~\cite{Martin:2005ch, Schofbeck:2006gs, Schofbeck:2007ib, Ibe:2012sx}. 
The mass difference between the lightest chargino and the lightest neutralino, $\delta_c=\big(m_{\tilde{\chi}_1^\pm}-m_{\tilde{\chi}_1^0}\big)/m_{\tilde{\chi}_1^0}$ becomes also very small if $\lvert\lvert\mu\rvert-M_2\rvert>>M_Z$. One can rewrite the LO $\tilde{\chi}_1^0\tilde{\chi}_1^0h$ and $\tilde{\chi}_1^0\tilde{\chi}_1^0H$ 
couplings for Wino-like LSP as~\cite{Hisano:2004pv},
\begin{align}
\mathcal{L}_{\tilde{\chi}_1^0\tilde{\chi}_1^0 h_i} = 
h_i \bar{\tilde{\chi}}_1^0  \Big[\mathbf{P_L} C_L^{\rm LO} + \mathbf{P_R} C_R^{\rm LO}\Big]\tilde{\chi}_{1}^0~.
\end{align}

For $h_i=h$,
\begin{align}
C_L^{\rm LO} = C_R^{\rm LO} = -\frac{g_2}{2} \frac{M_W}{M_2^2-\mu^2}\big(M_2+\mu s_{2\beta}\big)~,
\label{clcrlo1}
\end{align}

and for $h_i=H$,
\begin{align}
C_L^{\rm LO} = C_R^{\rm LO} = \frac{g_2}{2} \frac{M_W}{M_2^2-\mu^2}\mu c_{2\beta}~.
\label{clcrHlo2}
\end{align}

\textbf{$\bullet$ Effective neutralino-nucleon interactions\,:} We now present the effective Lagrangian governing the neutralino-nucleon scattering process and provide the corresponding formulae for the cross-section~\cite{Jungman:1995df, Bertone:2004pz, Goodman:1984dc, Griest:1988ma, Ellis:1987sh, Barbieri:1988zs, Drees:1993bu, Nath:1994ci, Ellis:2000ds, Vergados:2006sy, Oikonomou:2006mh, Ellis:2008hf, Ellis:2018dmb}. The effective interactions between non-relativistic neutralinos ($\tilde{\chi}_1^0$) and light quarks and gluon at the renormalization scale $\bar{\mu}_0 \simeq m_p$ can be represented as follows,
\begin{align}
\mathcal{L}^{\rm eff}=\sum_{q=u,d,s}\mathcal{L}_q^{\rm eff}+\mathcal{L}_g^{\rm eff},
\label{eq:L_eff}
\end{align}
where,
\begin{align}
\mathcal{L}_q^{\rm eff}&=\eta_q\bar{\tilde{\chi}}^0_1\gamma^\mu\gamma_5\tilde{\chi}^0_1\bar{q}\gamma_\mu\gamma_5 q + \lambda_qm_q\bar{\tilde{\chi}}^0_1\tilde{\chi}^0_1 \bar{q}q +\frac{g_q^{(1)}}{m_{\tilde{\chi}^0_1}}\bar{\tilde{\chi}}^0_1 i\partial^\mu\gamma^\nu\tilde{\chi}^0_1\mathcal{O}^q_{\mu\nu}+ \frac{g_q^{(2)}}{m^2_{\tilde{\chi}^0_1}}\bar{\tilde{\chi}}^0_1 (i\partial^\mu)(i\partial^\nu)\tilde{\chi}^0_1\mathcal{O}^q_{\mu\nu}~,\nonumber\\
\mathcal{L}_g^{\rm eff}&= \lambda_G\bar{\tilde{\chi}}^0_1\tilde{\chi}^0_1G^a_{\mu\nu}G^{a\,\mu\nu}+ \frac{g_G^{(1)}}{m_{\tilde{\chi}^0_1}}\bar{\tilde{\chi}}^0_1 i\partial^\mu\gamma^\nu\tilde{\chi}^0_1\mathcal{O}^g_{\mu\nu}+ \frac{g_G^{(2)}}{m^2_{\tilde{\chi}^0_1}}\bar{\tilde{\chi}}^0_1 (i\partial^\mu)(i\partial^\nu)\tilde{\chi}^0_1\mathcal{O}^g_{\mu\nu}~.
\label{eq:Lq_Lg}
\end{align}

The terms up to the second derivative of the neutralino field have been included above.
The spin-dependent interaction refers to the first term of $\mathcal{L}_q^{\rm eff}$ while the spin-independent {\it coherent} contributions arise from the second and the first term in the $\mathcal{L}_q^{\rm eff}$ and $\mathcal{L}_g^{\rm eff}$ respectively.
The third and fourth terms in $\mathcal{L}_q^{\rm eff}$, as well as the second and third terms in $\mathcal{L}_g^{\rm eff}$, are determined by the \textbf{twist-2} operators (traceless part of the energy-momentum tensor) for the quarks and gluons~\cite{Drees:1993bu, Hisano:2004pv}:

\begin{align}
    \mathcal{O}_{\mu\nu}^q \equiv &\frac{1}{2} \bar{q} i \Big[\partial_\mu\gamma_\nu + \partial_\nu\gamma_\mu -\frac{1}{2}g_{\mu\nu}\slashed{\partial}\Big]q~,\nonumber\\
    \mathcal{O}_{\mu\nu}^g \equiv &\Big[G^{a}_{\hspace{0.15cm}\mu}{}^{\rho} G^a_{\hspace{0.15cm}\rho\nu} +\frac{1}{4} g_{\mu\nu} G^a_{\alpha\beta}G^{a\,\alpha\beta}\Big]~,
\end{align}
where $G^a_{\alpha\beta}$ refers to gluon field strength tensor. Finally, the LSP scattering cross-section with target nuclei can be expressed in a compact form as,
\begin{align}
\sigma=\frac{4}{\pi}\left(\frac{m_{\tilde{\chi}^0_1}m_T}{m_{\tilde{\chi}^0_1}+m_T}\right)^2\Bigg[\{Zf_p+(A-Z)f_n\}^2+4\left(\frac{J+1}{J}\right)\{a_p\langle S_p\rangle + a_n\langle S_n\rangle\}^2\Bigg]~,
\label{eq:sigma_nucl}
\end{align} 
where $m_{\tilde{\chi}_1^0}$ and $m_T$ denote the mass of the LSP and the target nucleus, with $Z$ and $A$ representing the atomic and mass numbers of the target nucleus, respectively.
\vskip0.2cm
\textbf{SI contributions\,:} The spin-independent coupling of the neutralino with nucleon (of mass $m_N$), $f_N$ ($N=p,n$) in Eq.~\eqref{eq:sigma_nucl} can be expressed as
\begin{align}
\frac{f_N}{m_N}=&\sum_{q=u,d,s} \lambda_qf^{(N)}_q + \sum_{q=u,d,s,c,b} \bigg[\frac{3}{4}\bigl\{q\big(2,\bar{\mu}_0^2\big) + \bar{q}\big(2,\bar{\mu}_0^2\big)\bigr\}\big(g_q^{(1)} + g_q^{(2)}\big)(\bar{\mu}_0)\bigg] \nonumber\\
&-\frac{8\pi}{9\alpha_s}\lambda_Gf^{(N)}_{G} + \frac{3}{4}G\big(2,\bar{\mu}_0^2\big)\big(g_G^{(1)} + g_G^{(2)}\big)(\bar{\mu}_0)~,
\label{Eq:fnSI}
\end{align}
where the matrix elements of nucleon are defined as 
\begin{align}
f_q^{(N)} \equiv & \frac{1}{m_N}\langle N|m_q\bar{q}q|N\rangle~,\nonumber\\
f_G^{(N)} \equiv & 1-\sum_{u,d,s} f_q^{(N)}~,\nonumber\\
\langle N(p)|\mathcal{O}_{\mu\nu}^q|N(p)\rangle = &\frac{1}{m_N} \Big[p_\mu p_\nu -\frac{1}{4} m_N^2 g_{\mu\nu}\Big]\bigl\{q\big(2,\bar{\mu}_0^2\big) + \bar{q}\big(2,\bar{\mu}_0^2\big)\bigr\}~,\nonumber\\
\langle N(p)|\mathcal{O}_{\mu\nu}^g|N(p)\rangle = &\frac{1}{m_N} \Big[p_\mu p_\nu -\frac{1}{4} m_N^2 g_{\mu\nu}\Big] G\big(2,\bar{\mu}_0^2\big)~.
\end{align}
Here, $q\big(2,\bar{\mu}_0^2\big)$, $\bar{q}\big(2,\bar{\mu}_0^2\big)$, and $G\big(2,\bar{\mu}_0^2\big)$ are the second moments of quark, anti-quark and gluon distributions function, respectively,

\begin{align}
    q\big(2,\bar{\mu}_0^2\big) + \bar{q}\big(2,\bar{\mu}_0^2\big) = &\int_{0}^{1} dx\,\,x\big[q\big(x, \bar{\mu}_0^2\big) + \bar{q}\big(x, \bar{\mu}_0^2\big)\big]~,\nonumber\\
    G\big(2,\bar{\mu}_0^2\big) = &\int_0^1 dx\,\, x\,\, g\big(x, \bar{\mu}_0^2\big)~,
\end{align}
where the numerical values of the second moments can be found in Ref.~\cite{Hisano:2012wm, Ellis:2023ndh}.

The evaluation of $f_G^{(N)}$ in Eq.~\eqref{Eq:fnSI} involves effective interactions between the WIMP, heavy quarks, and gluons, which can be calculated using the trace anomaly of the energy-momentum tensor in QCD~\cite{SHIFMAN1978443, Drees:1993bu}. 
Here, one finds that heavy quark form factors are related to gluons.
\begin{align}
\langle N|m_Q\bar{Q}Q|N\rangle= & -\frac{\alpha_s}{12\pi} c_Q
\langle N|G^a_{\mu\nu}G^{a\mu\nu}|N\rangle~, \nonumber \\
m_N f_G^{(N)}= & -9 \frac{\alpha_s}{8\pi}\langle N|G^a_{\mu\nu}G^{a\mu\nu}|N\rangle,
\end{align}
with $\alpha_s=g_s^2/4\pi$ and the leading order
QCD correction $c_Q=1+11\alpha_s(m_Q)/4\pi$ is considered. 
The coefficient 
$\lambda_G$ in Eq.~\eqref{Eq:fnSI} is related to 
$\tilde\chi_1^0$-Higgs physics and heavy quarks. For the latter, one finds,
\begin{align}
\lambda_G \to -\frac{\alpha_s}{12\pi}\sum_{Q=c,b,t}c_Q\lambda_Q,
\end{align}
with $\lambda_Q$ can often be determined from $h_i Q\bar Q$ 
vertex at the tree level.

\section{An improved analysis of the SI-DD of Wino-like DM}
\label{sec:presentstatus}
In the subsequent subsections, we present a detailed analysis of the DD of Wino-like DM, specifically focusing on the improvement made in this work. As mentioned earlier, numerous studies have already explored the DD of the Wino-like DM.
We begin with a summary of the theoretical calculations of the SI-DD at LO and NLO, which are known in the literature. 

\subsection{Present status and our objective}
\label{subsec:presentstatus}
\begin{figure}[H]
	\centering
    \includegraphics[width=0.275\linewidth]{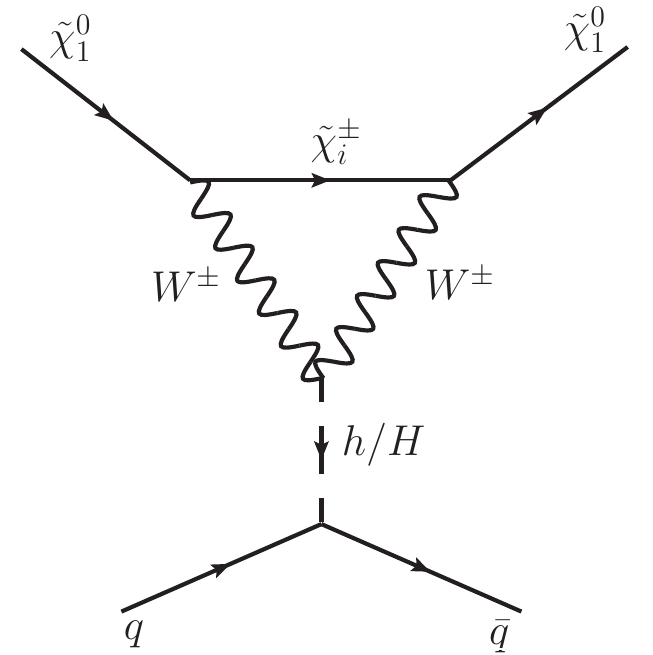}
	\hspace{0.7cm}\includegraphics[width=0.30\linewidth]{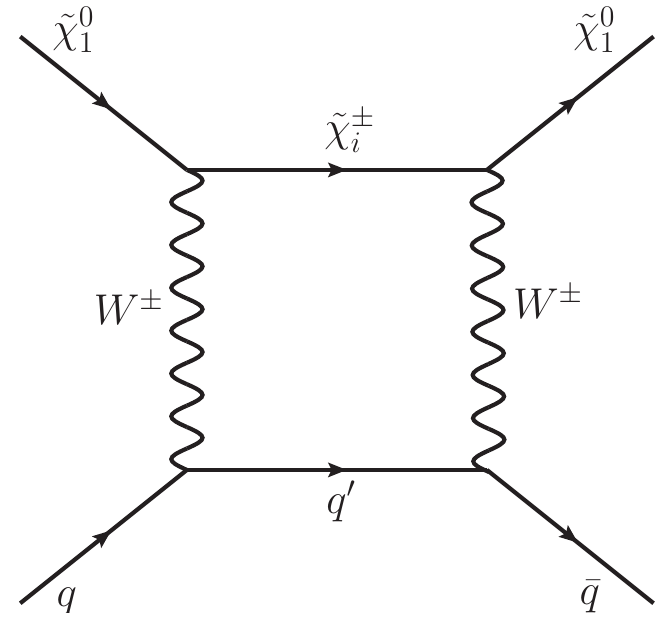}\\
    (a)\hspace{5.2cm}(b)
	\caption{One-loop triangle and box diagrams contributing to the neutralino-quark scattering. Similar diagrams replacing $W^\pm$ with $Z$ are found to be suppressed, assuming $\chi_1^0$ as a predominantly Wino-like state. The cross-diagram for the right figure is not shown.}
	\label{fig:boxdiag}
\end{figure}

\begin{figure}[H]
	\centering
    \includegraphics[width=0.252\linewidth]{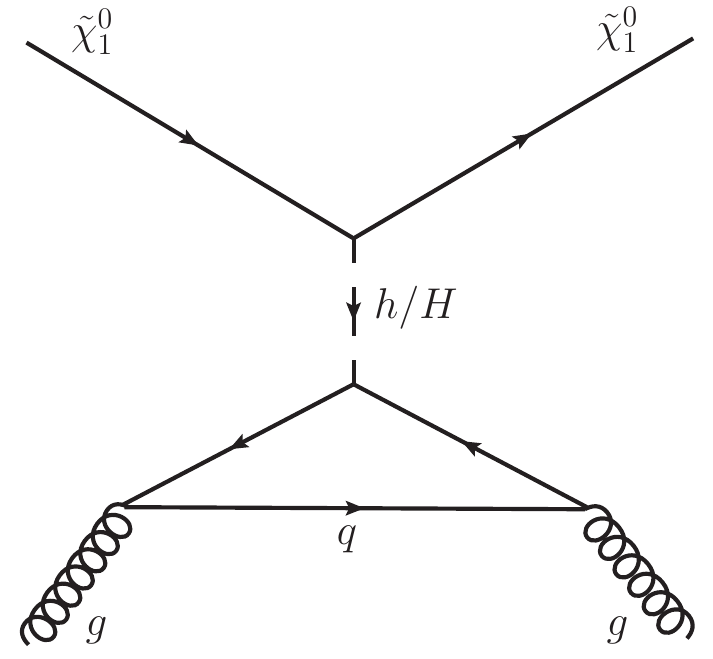}
	\includegraphics[width=0.252\linewidth]{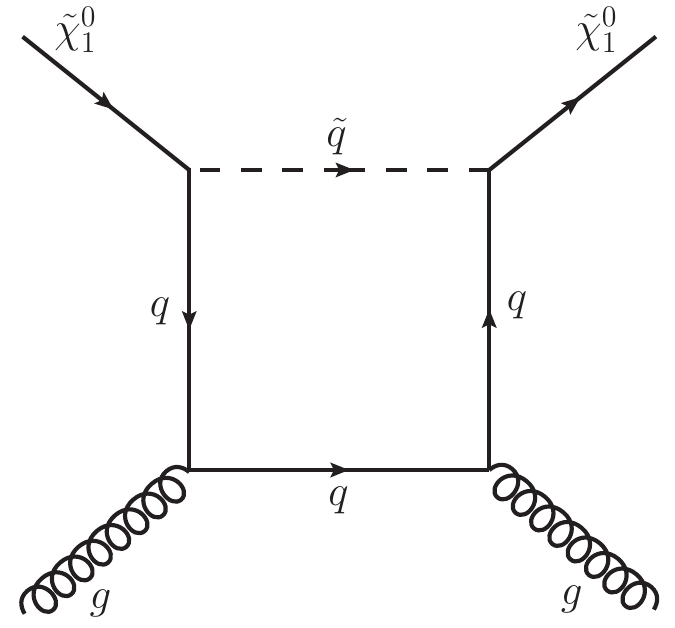}
    \includegraphics[width=0.215\linewidth]{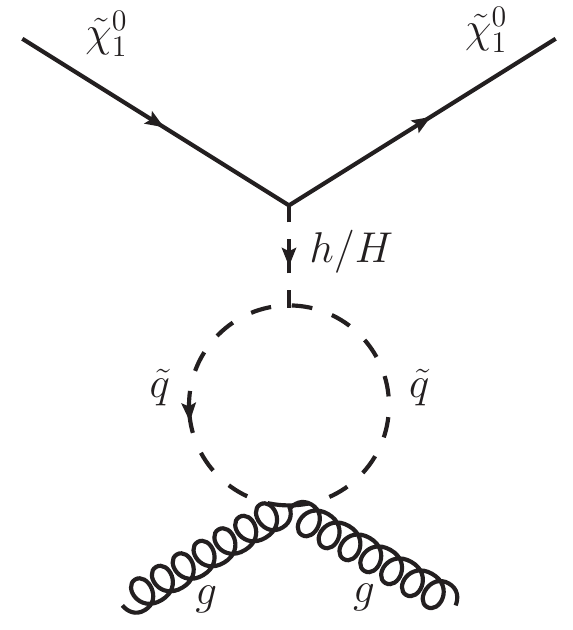}
	\includegraphics[width=0.255\linewidth]{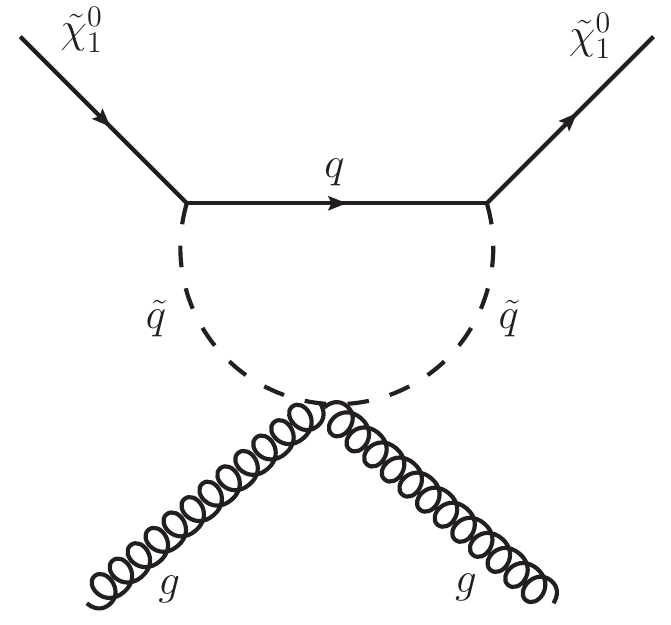}\\
    \hspace{-0.1cm}(a)\hspace{3.7cm}(b)\hspace{3.5cm}(c)\hspace{3.4cm}(d)
	\caption{One-loop diagrams contributing to the neutralino-gluon scattering. In figures (a) and (b), the quarks (squarks) can be replaced by squarks (quarks), resulting in additional diagrams that are not shown here. Additionally, the cross diagram for figure (b) is not shown.}
	\label{fig:boxdiag_sq}
\end{figure}

\begin{figure}[H]
	\centering
    \includegraphics[width=0.252\linewidth]{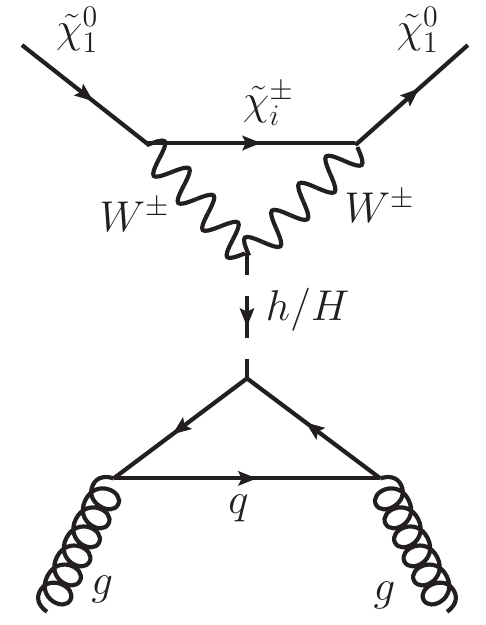}
	\hspace{0.7cm}\includegraphics[width=0.675\linewidth]{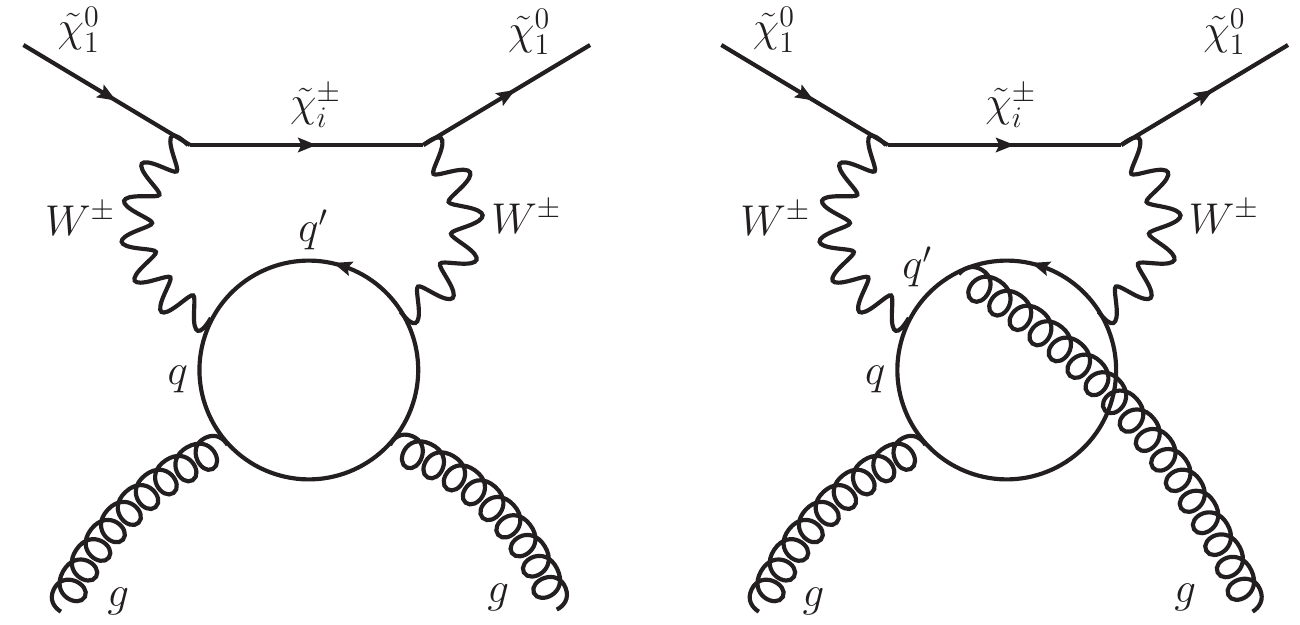}\\
     \hspace{-0.4cm}(a)\hspace{4.9cm}(b)\hspace{5.4cm}(c)
	\caption{Two-loop diagrams contributing to the neutralino-gluon scattering. Figure (a) is the reducible diagram, which depends on the $\tilde{\chi}_1^0\tilde{\chi}_1^0h_i$ vertex, whereas figures (b) and (c) are the irreducible ones.}
	\label{fig:twoloop}
\end{figure}

The LO neutralino-quark scattering is realized 
 through $t$-channel $h_i$ exchange while $\tilde \chi_1^0$-gluon 
 interactions appear at the one-loop involving heavy quarks and squarks. We may classify the beyond ``tree-level" contributions
as follows\,:
\begin{itemize}
\item \textbf{One-loop contributions\,:} The diagram in Fig.~\ref{fig:boxdiag}a contributes to the scalar quark operator through the coefficient $\lambda_q$ in Eq.~\eqref{eq:Lq_Lg}. Fig.~\ref{fig:boxdiag}b contributes to the quark \textbf{twist-2} operator through the coefficients $g_q^{(1)}$ and $g_q^{(2)}$ in Eq.~\eqref{eq:Lq_Lg}. The diagrams in Fig.~\ref{fig:boxdiag} are generally referred to as the NLO EW corrections to $\tilde\chi_1^0\tilde\chi_1^0 q \bar{q}$ process. However, in regard to LO
$\tilde\chi_1^0\tilde\chi_1^0 gg$ process, the diagrams in Fig.~\ref{fig:boxdiag_sq} contribute to the scalar gluon operator through the coefficient $\lambda_G$, while Fig.~\ref{fig:boxdiag_sq}b and Fig.~\ref{fig:boxdiag_sq}d additionally contribute to $g_G^{(1)}$ and $g_G^{(2)}$ in Eq.~\eqref{eq:Lq_Lg}~\cite{Drees:1992rr, Drees:1993bu}. All of these one-loop contributions (Fig.~\ref{fig:boxdiag}, \ref{fig:boxdiag_sq}) are considered in Ref.~\cite{Drees:1992rr, Drees:1993bu, Hisano:2010fy, Hisano:2012wm, Ellis:2023ndh, Hisano:2011cs, Hisano:2015rsa}.

Ref.~\cite{Essig:2007az, Cirelli:2005uq} studied the DD of DM in a model-independent way, considering the $SU(2)_L$ triplet fermion (equivalent to the Wino-like DM in the MSSM). They have considered only the scalar and quark \textbf{twist-2} contributions through the diagrams in Fig.~\ref{fig:boxdiag}.
    
\item \textbf{Two-loop contributions\,:} The two-loop or the NLO EW contributions to $\tilde \chi_1^0$-gluon scattering, depicted in Fig.~\ref{fig:twoloop}, contribute to 
$\lambda_G$ while Fig.~\ref{fig:twoloop}b and Fig.~\ref{fig:twoloop}c additionally
be counted in the gluon \textbf{twist-2} operators. The scalar gluon contributions are considered in Ref.~\cite{Hisano:2010fy, Hisano:2012wm, Ellis:2023ndh, Hisano:2011cs, Hisano:2015rsa}. The two-loop gluon \textbf{twist-2} contributions are very small and are neglected in all previous works.
\end{itemize}

\textbf{Our objective\,:} For a predominantly Wino-like DM ($|\mu|>>M_2$), 
the Higgsino compositions become completely insignificant; thus, the tree-level DM-DM-Higgs interaction becomes negligible. 
However, the NLO EW corrections shown in Fig.~\ref{fig:boxdiag} and \ref{fig:twoloop} turns out to be important, as these diagrams are not suppressed by the Higgsino fractions. Note that the diagrams presented in Fig.~\ref{fig:boxdiag_sq}b-\ref{fig:boxdiag_sq}d become relevant only when the squarks are not excessively heavy.
On the contrary, with a somewhat significant Higgsino composition, the Wino-like DM enjoys the \textit{not-so-small} tree-level interactions with the CP-even Higgs bosons of the MSSM. In this case, many more one-loop $\tilde{\chi}_1^0\tilde{\chi}_1^0h_i$ vertices along with the vertex counterterms for the renormalization may appear, which were not previously considered in the literature. 
Note that, our calculation assumes the most general scenario involving squarks and sleptons in the $\tilde{\chi}_1^0\tilde{\chi}_1^0h_i$ vertex at one-loop.
Thus, if one considers all the one-loop diagrams, shown in Fig.~\ref{fig:topology1}, incorporating all the EW particles of the MSSM together with all the existing contributions (via Fig.~\ref{fig:boxdiag}, \ref{fig:boxdiag_sq}, and \ref{fig:twoloop}), the DD cross-sections may be significantly changed, which we will discuss in the subsequent sections. In particular, assuming heavy squark masses (as preferred to comply with the LHC constraints), we will show that the EW SUSY particles in the MSSM can render significant changes to the SI-DD cross-section of $\chi_1^0$-nucleon scattering.

\subsection{Vertex corrections}
\label{subsec:ver}

\begin{figure}[H]
	\centering
	\includegraphics[width=1.0\linewidth]{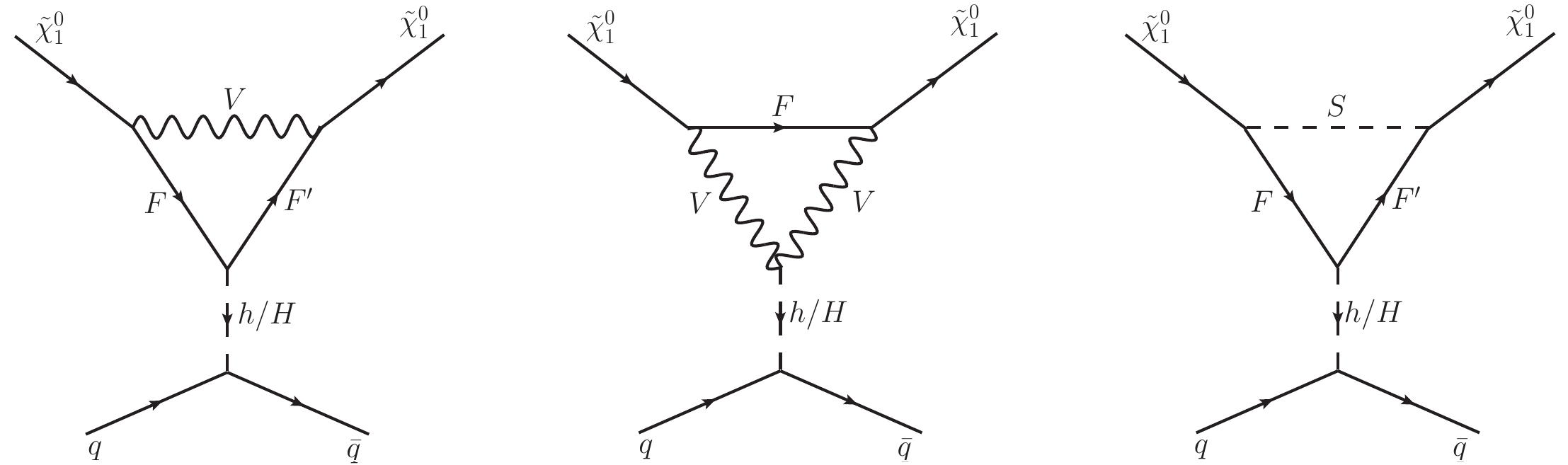}\\
    (a)\hspace{5.2cm}(b)\hspace{5.2cm}(c)
    \includegraphics[width=1.0\linewidth]{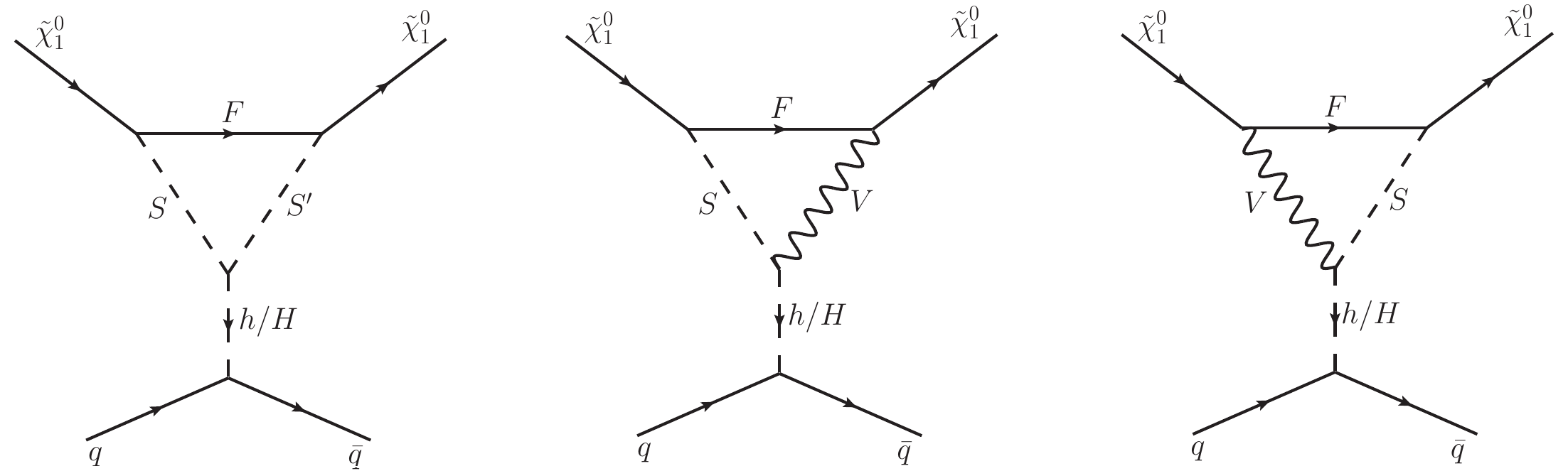}\\
    (d)\hspace{5.2cm}(e)\hspace{5.2cm}(f)
     \caption{One-loop triangular topologies contributing to the $\tilde{\chi}_1^0\tilde{\chi}_1^0h$ and $\tilde{\chi}_1^0\tilde{\chi}_1^0H$ vertices, which in turn give rise to contributions to the $\tilde{\chi}_1^0-q$ scattering process.
    Here $S,S^\prime \in \bigl\{h,\,\, H,\,\, A,\,\, H^\pm, \,\,G^0,\,\, G^\pm, \,\,\tilde{\ell}, \,\,\tilde{\nu}_\ell\,\, (\ell \in \{e, \mu, \tau\}), \,\,\tilde{q}\,\, (\tilde{q}= {\rm all\,\, the\,\, squarks})\bigr\}$; $F,F^\prime \in \bigl\{\tilde{\chi}_n^0, \,\,\tilde{\chi}_k^\pm \,\,(n = 1,...,4, \,\,{\rm and}\,\, k=1,2), \,\,\ell,\,\, \nu_\ell,\,\, q \,\,(q=u,c,t,d,s,b)\bigr\}$; $V = \bigl\{W^\pm,\,\, Z\bigr\}$.}
	\label{fig:topology1}
\end{figure}

Guided by our previous discussion, considering a Wino-like DM with some Higgsino components, alongside the diagram illustrated in Fig.~\ref{fig:boxdiag}a, the $\tilde{\chi}_1^0\tilde{\chi}_1^0h_i$ vertex undergoes radiative corrections from all other SM and SUSY particles.
Fig.~\ref{fig:topology1} illustrates all the triangular topologies contributing to the $\tilde{\chi}_1^0\tilde{\chi}_1^0h_i$ vertex corrections at one-loop level.

\noindent

 We use a general notation $S,S^\prime \in \bigl\{h,\,\, H,\,\, A,\,\, H^\pm, \,\,G^0,\,\, G^\pm, \,\,\tilde{\ell}, \,\,\tilde{\nu}_\ell\,\, (\ell \in \{e, \mu, \tau\}), \,\,\tilde{q}\,\, (\tilde{q}= {\rm all\,\, the\,\, squarks})\bigr\}$; $F,F^\prime \in \bigl\{\tilde{\chi}_n^0, \,\,\tilde{\chi}_k^\pm \,\,(n = 1,...,4, \,\,{\rm and}\,\, k=1,2), \,\,\ell,\,\, \nu_\ell,\,\, q \,\,(q=u,c,t,d,s,b)\bigr\}$; $V = \bigl\{W^\pm,\,\, Z\bigr\}$. For explicit calculations, we find a total of 612
diagrams where 306 diagrams for the $\tilde{\chi}_1^0\tilde{\chi}_1^0h$ vertex and 306 diagrams for the $\tilde{\chi}_1^0\tilde{\chi}_1^0H$ vertex at the particle level.

The total vertex corrections can be obtained as 
\begin{align}
\Gamma_{\tilde{\chi}_1^0\tilde{\chi}_1^0h_i} =& \Gamma_{\tilde{\chi}_1^0\tilde{\chi}_1^0h_i}^{(a)} + \Gamma_{\tilde{\chi}_1^0\tilde{\chi}_1^0h_i}^{(b)}+ \Gamma_{\tilde{\chi}_1^0\tilde{\chi}_1^0h_i}^{(c)}+
\Gamma_{\tilde{\chi}_1^0\tilde{\chi}_1^0h_i}^{(d)}+
\Gamma_{\tilde{\chi}_1^0\tilde{\chi}_1^0h_i}^{(e)}+
\Gamma_{\tilde{\chi}_1^0\tilde{\chi}_1^0h_i}^{(f)}\nonumber\\
=& C_L^{\rm 1L}\mathbf{P_L} + C_R^{\rm 1L} \mathbf{P_R}~,
\label{totalvertex:corrections}
\end{align}
 where $\Gamma_{\tilde{\chi}_1^0\tilde{\chi}_1^0h_i}^{(a),...,(f)}$ being the individual one-loop contributions from the diagrams \ref{fig:topology1}a,...,\ref{fig:topology1}f, respectively and $C^{\rm 1L}_{L, R}$ is the total one-loop corrections to the coefficients of the left- and right-handed projection operators in the $\tilde{\chi}_1^0\tilde{\chi}_1^0h_i$ vertex.

\subsection{Renormalization of the $\tilde{\chi}_1^0\tilde{\chi}_1^0h_i$ vertex\,: contributions of counterterms}
\label{subsec:renormalization}

In this section, we provide a brief overview of the various schemes utilized to renormalize the chargino and neutralino sectors within the MSSM framework, which is crucial for calculating the vertex counterterms of DM-DM-Higgs interactions. For detailed construction on the renormalization process, including counterterms and renormalization constants, readers are directed to Ref.~\cite{Eberl:2001eu, Fritzsche:2002bi, Oller:2003ge, Oller:2005xg, Drees:2006um, Fowler:2009ay, Heinemeyer:2011gk, Chatterjee:2012hkk, Bharucha:2012re, Hahn:2015ghv}.
Within this framework, the SUSY parameters defining neutral and charged fermions consist of the EW gaugino mass parameters $M_1$, $M_2$, and the supersymmetric Higgsino mass parameter $\mu$. The mass matrices incorporate the masses of the EW gauge bosons with mixing angle $\theta_W$ and $\tan\beta$; all these parameters are renormalized independently from the chargino and neutralino sectors.
The details of implementation are discussed in Ref.~\cite{Fritzsche:2013fta}, which also covers the Feynman rules for counterterms in a general complex MSSM (cMSSM).

The Fourier-transformed MSSM Lagrangian, which is bilinear in the chargino and neutralino fields, can be expressed as~\cite{Eberl:2001eu,Fritzsche:2002bi, 
Oller:2003ge, Oller:2005xg,Drees:2006um, Fowler:2009ay,Heinemeyer:2011gk,
Chatterjee:2012hkk,Bharucha:2012re, Hahn:2015ghv},
\begin{align}
\mathcal{L}_{\tilde{\chi}^{\pm}\tilde{\chi}^0} &= \bar{\tilde{\chi}}_i^{\pm}\slashed{p}\mathbf{P_L}\tilde{\chi}_i^{\pm} + \bar{\tilde{\chi}}_i^{\pm}\slashed{p}\mathbf{P_R}\tilde{\chi}_i^{\pm} - \bar{\tilde{\chi}}_i^{\pm}\big[\mathbb{V}^{*}\overline{\mathbb{M}}_{\tilde{\chi}^\pm}^{\rm T}\mathbb{U}^{\dagger}\big]_{ij}\mathbf{P_L}\tilde{\chi}_j^{\pm} - \bar{\tilde{\chi}}_i^{\pm}\big[\mathbb{U}\overline{\mathbb{M}}_{\tilde{\chi}^\pm}^{*}\mathbb{V}^{\rm T}\big]_{ij}\mathbf{P_R}\tilde{\chi}_j^{\pm}\nonumber\\
&+\frac{1}{2}\Big(\bar{\tilde{\chi}}_m^{0}\slashed{p}\mathbf{P_L}\tilde{\chi}_m^{0} + \bar{\tilde{\chi}}_m^{0}\slashed{p}\mathbf{P_R}\tilde{\chi}_m^{0} -\bar{\tilde{\chi}}_m^{0}\big[\mathbb{N}^{*}\overline{\mathbb{M}}_{\tilde{\chi}^0}\mathbb{N}^{\dagger}\big]_{mn}\mathbf{P_L}\tilde{\chi}_n^{0}-\bar{\tilde{\chi}}_m^{0}\big[\mathbb{N}\overline{\mathbb{M}}_{\tilde{\chi}^0}^{*}\mathbb{N}^{\rm T}\big]_{mn}\mathbf{P_R}\tilde{\chi}_n^{0}\Big)~,
\end{align}
where $i,j=1,2$,~$m,n=1,...,4$. We recall that, $\mathbb{U}$, $\mathbb{V}$ and $\mathbb{N}$ diagonalize the chargino and neutralino mass matrices $\overline{\mathbb{M}}_{\tilde{\chi}^\pm}$ and $\overline{\mathbb{M}}_{\tilde{\chi}^0}$ in the weak basis, respectively~\cite{Bisal:2023iip}.

We note the following replacements made to the parameters and fields\,:
\begin{align}
M_1&\rightarrow M_1 + \delta M_1~,\\
M_2&\rightarrow M_2 + \delta M_2~,\\
\mu &\rightarrow \mu + \delta \mu~,\\
\mathbf{P_L}\tilde{\chi}_i^{\pm} &\rightarrow \Bigg[\mathds{1} + \frac{1}{2}\delta\mathbb{Z}_{\tilde{\chi}^{\pm}}^{L}\Bigg]_{ij}\mathbf{P_L}\tilde{\chi}_{j}^{\pm}~,\\
\mathbf{P_R}\tilde{\chi}_i^{\pm} &\rightarrow \Bigg[\mathds{1} + \frac{1}{2}\delta\mathbb{Z}_{\tilde{\chi}^{\pm}}^{R}\Bigg]_{ij}\mathbf{P_R}\tilde{\chi}_{j}^{\pm}~,
\end{align}

\begin{align}
\mathbf{P_L}\tilde{\chi}_m^{0} &\rightarrow \Bigg[\mathds{1} + \frac{1}{2}\delta\mathbb{Z}_{\tilde{\chi}^{0}}\Bigg]_{mn}\mathbf{P_L}\tilde{\chi}_{n}^{0}~,\\
\mathbf{P_R}\tilde{\chi}_m^{0} &\rightarrow \Bigg[\mathds{1} + \frac{1}{2}\delta\mathbb{Z}_{\tilde{\chi}^{0}}^{*}\Bigg]_{mn}\mathbf{P_R}\tilde{\chi}_{n}^{0}~,
\end{align} 
where 
$\delta\mathbb{Z}_{\tilde{\chi}^{\pm},\tilde{\chi}^{0}}$ refers to field renormalization constants for the physical states, is in general $2\times 2$ or $4\times 4$ matrices respectively. The parameter counterterms are generally complex; we need two renormalization conditions to fix those counterterms (one for the real part and another for the complex part). Given the fact that the transformation matrices are not renormalized, we may write,
\begin{align}
\overline{\mathbb{M}}_{\tilde{\chi}^\pm} \rightarrow&\, \overline{\mathbb{M}}_{\tilde{\chi}^\pm} + \delta\overline{\mathbb{M}}_{\tilde{\chi}^\pm}~,\\
\overline{\mathbb{M}}_{\tilde{\chi}^0} \rightarrow& \,\overline{\mathbb{M}}_{\tilde{\chi}^0} + \delta\overline{\mathbb{M}}_{\tilde{\chi}^0}~,
\end{align}
with
\begin{align}
\delta\overline{\mathbb{M}}_{\tilde{\chi}^\pm}=\left(\begin{array}{c c }
\delta M_2 & \sqrt{2}\,\delta(M_W s_\beta)\\
\sqrt{2}\,\delta(M_W c_\beta) & \delta\mu\\
\end{array}\right)~,
\label{eq:Counter_chargino_mass_matrix}
\end{align}
and 
\begin{align}
\delta\overline{\mathbb{M}}_{\tilde{\chi}^0}=\left(\begin{array}{c c c c}
\delta M_1 & 0 & -\delta(M_Zs_W c_\beta) & \delta(M_Zs_W s_\beta) \\
0 & \delta M_2 & \delta(M_Zc_W c_\beta) & -\delta(M_Zc_W s_\beta)\\
-\delta(M_Zs_W c_\beta) & \delta(M_Zc_W c_\beta) & 0 & -\delta\mu\\
\delta(M_Zs_W s_\beta) & -\delta(M_Zc_W s_\beta) & -\delta\mu & 0\\
\end{array}\right).
\label{eq:Counter_nutralino_mass_matrix}
\end{align}

Similarly, for the diagonalized matrices $\mathbb{M}_{\tilde{\chi}^{\pm}}$ and $\mathbb{M}_{\tilde{\chi}^{0}}$, one can write
\begin{align}
\mathbb{M}_{\tilde{\chi}^{\pm}}&\rightarrow \mathbb{M}_{\tilde{\chi}^{\pm}} + \delta \mathbb{M}_{\tilde{\chi}^{\pm}} = \mathbb{M}_{\tilde{\chi}^{\pm}} + \mathbb{V}^{*}\delta\overline{\mathbb{M}}_{\tilde{\chi}^\pm}^{\rm T} \mathbb{U}^{\dagger}~,\\
\mathbb{M}_{\tilde{\chi}^{0}}&\rightarrow \mathbb{M}_{\tilde{\chi}^{0}} + \delta \mathbb{M}_{\tilde{\chi}^{0}} = \mathbb{M}_{\tilde{\chi}^{0}} + \mathbb{N}^{*}\delta\overline{\mathbb{M}}_{\tilde{\chi}^0}^{\rm T} \mathbb{N}^{\dagger}~.
\label{eq:chaneucount}
\end{align}

We can decompose the self energies into left- and right-handed vector and scalar coefficients in the following way\,:
\begin{align}
\big[\Sigma_{\tilde{\chi}}(p^2)\big]_{\ell m} = \slashed{p}\mathbf{P_L} \big[\Sigma_{\tilde{\chi}}^L(p^2)\big]_{\ell m} + \slashed{p}\mathbf{P_R} \big[\Sigma_{\tilde{\chi}}^R(p^2)\big]_{\ell m} + \mathbf{P_L} \big[\Sigma_{\tilde{\chi}}^{SL}(p^2)\big]_{\ell m} + \mathbf{P_R}\big[\Sigma_{\tilde{\chi}}^{SR}(p^2)\big]_{\ell m}~.
\label{Self_energy}
\end{align}

The expressions for renormalized self-energies can be located in Ref.~\cite{Eberl:2001eu, Fritzsche:2002bi, Oller:2003ge, Oller:2005xg, Drees:2006um, Fowler:2009ay, Heinemeyer:2011gk, Chatterjee:2012hkk, Bharucha:2012re, Fritzsche:2013fta}. The above expressions are utilized to compute the counterterms $\delta M_1$, $\delta M_2$, and $\delta \mu$, ensuring that the masses of $\tilde{\chi}_{1,2}^\pm$ and $\tilde{\chi}_n^0$ ($n=1,...,4$) correspond to the poles of their respective propagators. This approach, termed $\mathtt{CCN}[n]$, involves selecting $\tilde{\chi}_n^0$ 
as on-shell, with ``$\mathtt{C}$" representing chargino, ``$\mathtt{N}$" neutralino, and ``$n$" indicating the on-shell neutralino.
For instance, the $\mathtt{CCN[1]}$ scheme requires the mass of the primarily Bino-like lightest neutralino to be chosen on-shell for numerical stability~\cite{Chatterjee:2011wc}, while non-Bino-like LSP scenarios may exhibit large unphysical contributions if the LSP is taken as on-shell~\cite{Baro_2009}. This scheme accommodates Bino-dominated mixed LSP scenarios, including Bino-Higgsino or even Bino-Wino-Higgsino neutralinos. However, for other hierarchical mass patterns, such as $\lvert M_2\rvert< \lvert M_1\rvert,|\mu|$ or $|\mu|< \lvert M_1\rvert, \lvert M_2\rvert$, the $\mathtt{CCN}$[1] scheme may yield unstable results, necessitating alternative schemes like $\mathtt{CCN}$[2], $\mathtt{CCN}$[3] or $\mathtt{CCN}$[4]~\cite{Heinemeyer:2023pcc, Bharucha:2012re}. 
We focus on the mass patterns $M_2<M_1<\lvert\mu\rvert$ and $M_2<\lvert\mu\rvert<M_1$,
where the LSP becomes Wino-like with some Higgsino components. In the $\mathtt{CCN}$ scheme, since the Bino-like state must be taken on-shell to achieve numerically stable results, we use $\mathtt{CCN}$[2] scheme for the scenario $M_2<M_1<\lvert\mu\rvert$ and $\mathtt{CCN}$[4] scheme for the scenario $M_2<\lvert\mu\rvert<M_1$.
On the other hand, in the ``$\mathtt{CNN}$" scheme, one of the two charginos and two neutralinos $\tilde{\chi}_\ell^0$ and $\tilde{\chi}_m^0$ are taken to be on-shell~\cite{Drees:2006um, Chatterjee:2011wc, Heinemeyer:2023pcc}. As we are interested in Wino-like LSP scenarios, we adhere to imposing on-shell conditions for the two charginos and one neutralino.

\begin{figure}[H]
	\centering
	\includegraphics[width=0.65\linewidth]{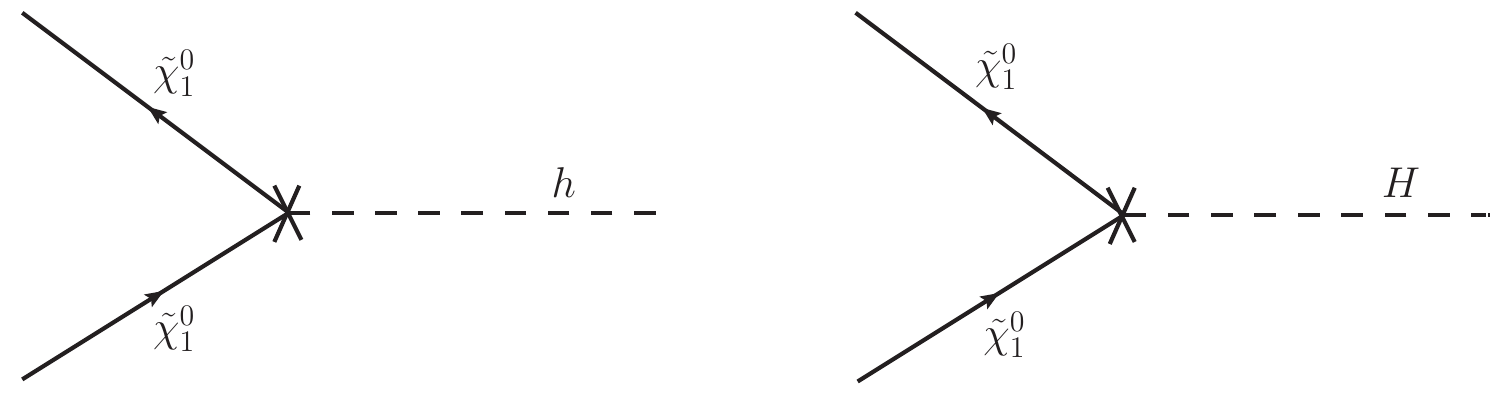}\\
    (a)\qquad\qquad\qquad\qquad\qquad\qquad\qquad\qquad\qquad\quad(b)
    \caption{ The counterterm diagrams for the $\tilde{\chi}^0_1 \tilde{\chi}^0_1 h$ and $\tilde{\chi}^0_1 \tilde{\chi}^0_1 H$ vertices.}
	\label{fig:chichihicounter}
\end{figure}

To achieve UV-finite results, counterterm diagrams for the $\tilde{\chi}^0_1 \tilde{\chi}^0_1 h$ and $\tilde{\chi}^0_1 \tilde{\chi}^0_1 H$ vertices need to be incorporated alongside the one-loop corrected diagrams. The field renormalization constants mentioned above can be used to compute the vertex counterterms. Ultimately, the expression for the vertex counterterm can be written as follows (see Fig.~\ref{fig:chichihicounter}a and Fig.~\ref{fig:chichihicounter}b)\,:
\begin{align}
\delta \Gamma_{\tilde{\chi}_1^0\tilde{\chi}_1^0h_i} = \mathbf{P_L} \delta C_L(h_i) + \mathbf{P_R} \delta C_R(h_i)~,
\label{countyerterm:eqngamma}
\end{align}
where $\delta C_{L,R}(h_i)$ for the SM-like Higgs can be written as
\begin{align}
\delta C_L(h) =& -\frac{e}{4c_Ws_W^2}\Bigg[\frac{4}{c_W^2}\bigl\{\big(c_W^2\delta Z_e+s_W\delta s_W\big)s_W^2\mathcal{N}^{*}_{11} + c_W c_W^2 \big(\delta s_W - s_W\delta Z_e\big)\mathcal{N}^{*}_{12}\bigr\}\big(s_\alpha \mathcal{N}^{*}_{13}+c_\alpha\mathcal{N}^{*}_{14}\big) \nonumber\\
& + s_W\Bigl\{2\big(s_W \mathcal{N}^{*}_{11} - c_W\mathcal{N}^{*}_{12}\big)\bigl\{\big(2\big[\delta\mathbf{Z}^L_{\tilde{\chi}^0}\big]_{11} + \delta Z_{hh}\big)\big(s_\alpha\mathcal{N}^{*}_{13}+c_\alpha\mathcal{N}^{*}_{14}\big) - \delta Z_{hH}\big(c_\alpha \mathcal{N}^{*}_{13}-s_\alpha\mathcal{N}^{*}_{14}\big)\bigr\}\nonumber\\
&+ \big(\big[\delta\mathbf{Z}^L_{\tilde{\chi}^0}\big]_{12} + \big[\delta\mathbf{Z}^L_{\tilde{\chi}^0}\big]_{21}\big)\bigl\{\big(s_\alpha \mathcal{N}^{*}_{13}+c_\alpha\mathcal{N}^{*}_{14}\big)\big(s_W\mathcal{N}^{*}_{21}-c_W\mathcal{N}^{*}_{22}\big)+\big(s_W\mathcal{N}^{*}_{11}-c_W\mathcal{N}^{*}_{12}\big)\big(s_\alpha\mathcal{N}^{*}_{23}\nonumber\\
&+c_\alpha\mathcal{N}^{*}_{24}\big)\bigr\} + \big(\big[\delta\mathbf{Z}^L_{\tilde{\chi}^0}\big]_{13} + \big[\delta\mathbf{Z}^L_{\tilde{\chi}^0}\big]_{31}\big)\bigl\{\big(s_\alpha\mathcal{N}^{*}_{13}+ c_\alpha\mathcal{N}^{*}_{14}\big)\big(s_W\mathcal{N}^{*}_{31}-c_W\mathcal{N}^{*}_{32}\big) + \big(s_W\mathcal{N}^{*}_{11}-c_W\mathcal{N}^{*}_{12}\big)\nonumber\\
&\times\big(s_\alpha\mathcal{N}^{*}_{33}+c_\alpha\mathcal{N}^{*}_{34}\big)\bigr\} + \big(\big[\delta\mathbf{Z}^L_{\tilde{\chi}^0}\big]_{14} + \big[\delta\mathbf{Z}^L_{\tilde{\chi}^0}\big]_{41}\big)\bigl\{\big(s_\alpha\mathcal{N}^{*}_{13}+c_\alpha\mathcal{N}^{*}_{14}\big)\big(s_W\mathcal{N}^{*}_{41}-c_W\mathcal{N}^{*}_{42}\big) + \big(s_W\mathcal{N}^{*}_{11}\nonumber\\
& -c_W\mathcal{N}^{*}_{12}\big)\big(s_\alpha\mathcal{N}^{*}_{43}+c_\alpha\mathcal{N}^{*}_{44}\big)\bigr\}\Bigr\}\Bigg]
\end{align}
and 
\begin{align}
\delta C_R(h) =& -\frac{e}{4c_W s_W^2}\Bigg[\frac{4}{c_W^2}\bigl\{\big(c_W^2\delta Z_e+s_W\delta s_W\big)s_W^2\mathcal{N}_{11} + c_W c_W^2 \big(\delta s_W - s_W\delta Z_e\big)\mathcal{N}_{12}\bigr\}\big(s_\alpha \mathcal{N}_{13}+c_\alpha\mathcal{N}_{14}\big) \nonumber\\
& +s_W\Bigl\{2\big(s_W\mathcal{N}_{11}-c_W\mathcal{N}_{12}\big)\bigl\{\big(s_\alpha\delta Z_{hh} - c_\alpha\delta Z_{hH}\big)\mathcal{N}_{13} + \big(c_\alpha\delta Z_{hh} + s_\alpha\delta Z_{hH}\big)\mathcal{N}_{14} + \big(2\big[\delta \mathbf{Z}_{\tilde{\chi}^0}^{R}\big]_{11}\big)\nonumber\\
&\times \big(s_\alpha\mathcal{N}_{13}+c_\alpha\mathcal{N}_{14}\big)\bigr\} + \big(\big[\delta\mathbf{Z}_{\tilde{\chi}^0}^{R}\big]_{12} + \big[\delta\mathbf{Z}_{\tilde{\chi}^0}^{R}\big]_{21}\big) \bigl\{\big(s_\alpha\mathcal{N}_{13}+c_\alpha\mathcal{N}_{14}\big)\big(s_W\mathcal{N}_{21}-c_W\mathcal{N}_{22}\big) + \big(s_W\mathcal{N}_{11}\nonumber\\
&-c_W\mathcal{N}_{12}\big)\big(s_\alpha\mathcal{N}_{23}+ c_\alpha \mathcal{N}_{24}\big)\bigr\} + \big(\big[\delta\mathbf{Z}_{\tilde{\chi}^0}^R\big]_{13} + \big[\delta\mathbf{Z}_{\tilde{\chi}^0}^R\big]_{31}\big)\bigl\{\big(s_\alpha\mathcal{N}_{13} + c_\alpha\mathcal{N}_{14}\big)\big(s_W\mathcal{N}_{31} - c_W\mathcal{N}_{32}\big) \nonumber\\
&+ \big(s_W\mathcal{N}_{11}-c_W\mathcal{N}_{12}\big)\big(s_\alpha\mathcal{N}_{33}+c_\alpha\mathcal{N}_{34}\big)\bigr\} + \big(\big[\delta\mathbf{Z}_{\tilde{\chi}^0}^R\big]_{14} + \big[\delta\mathbf{Z}_{\tilde{\chi}^0}^R\big]_{41}\big)\bigl\{\big(s_\alpha\mathcal{N}_{13}+c_\alpha\mathcal{N}_{14}\big)\big(s_W\mathcal{N}_{41}\nonumber\\
&-c_W\mathcal{N}_{42}\big) + \big(s_W\mathcal{N}_{11}-c_W\mathcal{N}_{12}\big)\big(s_\alpha\mathcal{N}_{43} + c_\alpha\mathcal{N}_{44}\big)\bigr\}\Bigr\} \Bigg]~,
\end{align}
where $\mathcal{N}_{ij}$ are the elements of the $4\times4$ unitary matrix that diagonalizes 
$\overline{\mathbb{M}}_{\tilde{\chi}^0}$. 
For the renormalization constants related to the Higgs sector \big(e.g., involving $\delta Z_{hh}$, $\delta Z_{hH}$, and $\delta Z_{HH}$\big) and
the SM \big(e.g., involving $\delta Z_{e}$ and $\delta s_{W}$\big), we refer to Ref.~\cite{Bisal:2023fgb, Bisal:2023iip}. Furthermore, the expressions for $\delta\mathbf{Z}_{\tilde{\chi}_1^0}^{L, R}$, which involve renormalized self-energies and counterterms of the mass matrices of the physical states can be found in Ref.~\cite{Heinemeyer:2011gk, Bharucha:2012re}. 
Similarly, the counterterm for the heavy Higgs can be obtained by the replacements $s_\alpha\to c_\alpha$, $c_\alpha\to -s_\alpha$, $\delta Z_{hh}\to \delta Z_{HH}$, and $\delta Z_{hH}\to -\delta Z_{hH}$.
We use the fine-structure constant $\alpha=\alpha(0)=1/137.0359996$ defined at the Thomson limit.~\footnote{$\mathtt{FormCalc}$ computes the charge renormalization constant at the Thomson limit, i.e., $\delta Z_e^{(0)}$~\cite{Fritzsche:2013fta}, and we use the same prescription for $\alpha(0)$ or $e(0)$. 
For other choices, see Ref.~\cite{Chatterjee:2012hkk, Denner:1991kt, Hagiwara:2011af, Steinhauser:1998rq}. }
To the end, we simply add the vertex corrections and counterterms as $\Gamma_{\tilde{\chi}_1^0\tilde{\chi}_1^0h_i} + \delta \Gamma_{\tilde{\chi}_1^0\tilde{\chi}_1^0h_i}$ to
obtain the UV-finite amplitude where $\Gamma_{\tilde{\chi}_1^0\tilde{\chi}_1^0 h_i}$ and $\delta \Gamma_{\tilde{\chi}_1^0\tilde{\chi}_1^0h_i}$ are defined in  Eq.~\eqref{totalvertex:corrections} and Eq.~\eqref{countyerterm:eqngamma}, respectively.

Note that, as usual, we define the on-shell (physical) masses as the poles of the real parts of the one-loop corrected propagators. Therefore, the on-shell chargino and neutralino masses can be expressed as~\cite{Chatterjee:2011wc, Eberl:2001eu, Fritzsche:2002bi, Oller:2003ge}
\begin{align}
    m_{\tilde{\chi}_i^\pm}^{\rm OS} = m_{\tilde{\chi}_i^\pm} + \Big[\mathbb{U}^{*} \delta \overline{\mathbb{M}}_{\tilde{\chi}_i^\pm} \mathbb{V}^{-1}\Big]_{ii} - \delta m_{\tilde{\chi}_i^\pm}~,
    \label{eq:onshellchar}
\end{align}
\begin{align}
    m_{\tilde{\chi}_i^0}^{\rm OS} = m_{\tilde{\chi}_i^0} + \Big[\mathbb{N}^{*} \delta \overline{\mathbb{M}}_{\tilde{\chi}_i^0} \mathbb{N}^{-1}\Big]_{ii} - \delta m_{\tilde{\chi}_i^0}~.
    \label{eq:onshellneu}
\end{align}
Here, $m_{\tilde{\chi}_i^\pm}$ and $m_{\tilde{\chi}_i^0}$ denote the tree-level masses for the charginos and neutralinos. The expression for $\delta m_f$ ($f=\tilde{\chi}_i^\pm, \tilde{\chi}_i^0$) is obtained as 
\begin{align}
    \delta m_f = \frac{1}{2} m_f \Big[\widetilde{\rm Re}\Sigma_{f}^{L}(m_f^2) + \widetilde{\rm Re}\Sigma_{f}^{R}(m_f^2)\Big] + \frac{1}{2} \Big[\widetilde{\rm Re}\Sigma_{f}^{SL}(m_f^2) + \widetilde{\rm Re}\Sigma_{f}^{SR}(m_f^2)\Big]~,
\end{align}
where $\widetilde{\rm Re}$ takes the real parts of the loop integrals while considering both the real and imaginary parts of the complex couplings.
 
Since we are using on-shell scheme, we always have $m_{\tilde{\chi}_1^\pm}^{\rm OS} = m_{\tilde{\chi}_1^\pm}$.
Consequently, the mass difference between the lightest chargino ($\tilde{\chi}_1^\pm$) and the lightest neutralino ($\tilde{\chi}_1^0$) can be expressed using Eq.~\eqref{eq:onshellchar} and \eqref{eq:onshellneu} as
\begin{align}
    \Delta m(\tilde{\chi}_1^\pm, \tilde{\chi}_1^0) &= m_{\tilde{\chi}_1^\pm}^{\rm OS} - m_{\tilde{\chi}_1^0}^{\rm OS}\nonumber\\
    &= \big(m_{\tilde{\chi}_1^\pm} - m_{\tilde{\chi}_1^0}\big) - \Big[\mathbb{N}^{*} \delta \overline{\mathbb{M}}_{\tilde{\chi}_1^0} \mathbb{N}^{-1}\Big]_{11} + \delta m_{\tilde{\chi}_1^0}~.
    \label{eq:onshellsplitting}
\end{align}

In the computations of the neutralino-nucleon scattering, the external Higgs bosons ($h$ and $H$) in the $\tilde{\chi}_1^0\tilde{\chi}_1^0h_i$ vertex would be considered off-shell, with a four-momentum denoted as $k$, which is known as the momentum transfer. 
The momentum transfer $k$ is generally very small \big($k^2\sim \mathcal{O}(10^{-6})$ $\rm GeV^2$ for $v_{\tilde{\chi}_1^0} \sim 10^{-3}$\big) for the elastic scattering process. It may be noted here that $k^2\sim 0$ is assumed for the numerical evaluation.

\section{Improvements done within \texorpdfstring{$\mathtt{MicrOMEGAs}$}{MicrOMEGAs}}
\label{sec:improvementMicro}
Here we discuss the chronology of implementation of vertex corrections within \texorpdfstring{$\mathtt{MicrOMEGAs}$}{MicrOMEGAs} starting from the loop processes along with the validation of the code with the existing
literature. \\ \vskip 0.2 cm
\textbf{Implementation\,:} Based on the discussion in the earlier section, we summarize here the necessary steps to evaluate the renormalized $\tilde{\chi}_1^0\tilde{\chi}_1^0h_i$ vertex and the improved SI-DD cross-section at NLO. We use $\mathtt{FeynArts}$-3.11~\cite{Hahn:2000kx, KUBLBECK1990165, Hahn:2001rv, Fritzsche:2013fta}, $\mathtt{FormCalc}$-9.9~\cite{Hahn:1998yk, Fritzsche:2013fta}, $\mathtt{LoopTools}$-2.16~\cite{Hahn:1998yk}, $\mathtt{SARAH}$-4.14.5~\cite{Staub:2017jnp, Staub:2013tta, Staub:2015kfa}, $\mathtt{SPheno}$-4.0.4~\cite{Porod:2003um, Staub:2017jnp}, and $\mathtt{MicrOMEGAs}$-5.0.4~\cite{Belanger:2001fz,Belanger:2006is,Belanger:2008sj,Belanger:2013oya} (for a recent tool related to DM-nucleon NLO cross-section, see~\cite{Harz:2023llw}) at different stages of the computations as discussed below.

\begin{itemize}
    \item First, we generate all the one-loop and counterterm Feynman diagrams for the $\tilde{\chi}_1^0\tilde{\chi}_1^0h_i$ vertex and create the amplitude in the Feynman gauge using $\mathtt{FeynArts}$.

    \item Then, we use $\mathtt{FormCalc}$ to evaluate the loop integrals over the internal momenta and write the amplitude in terms of the Passarino-Veltman (PV) scalar integrals. 
    
    \item We evaluate all the renormalization constants by adopting the $\mathtt{CCN}[2]$ and $\mathtt{CCN}[4]$ schemes using $\mathtt{FormCalc}$, followed by the calculation of the amplitude for the vertex counterterms. 

    \item Then, we export the whole analytical expressions for the vertices and the corresponding counterterms into different $\mathtt{Fortran}$ subroutines. 

    \item For the numerical values of the MSSM parameters, we use the spectrum calculator $\mathtt{SPheno}$, which uses the $\mathtt{SARAH}$ generated model file for the MSSM. 

    \item We evaluate the numerical values for the vertices and the counterterms using $\mathtt{LoopTools}$, which uses the inputs from the output of $\mathtt{SPheno}$. We find that all the UV-divergencies cancel and obtain a finite result for the $\tilde{\chi}_1^0\tilde{\chi}_1^0h_i$ vertex at NLO.
The complete NLO vertex includes $C^{\rm LO}_{L, R}$, one-loop vertex corrections $C^{1\rm L}_{L, R}$, and contributions from the counterterms $\delta C_{L, R}$ as 
\begin{align}
    C^{\rm NLO}_{L,R}= C^{\rm LO}_{L, R} + C^{\rm 1L}_{L, R} + \delta C_{L, R}. 
    \label{eqn:nlo}
\end{align}    
The NLO vertices $C^{\rm NLO}_{L}= C^{\rm NLO}_{R}$ like the LO ones. Note that we use tree-level masses for all the particles appearing in the loop to get the
UV-finite result (see e.g.,~\cite{Bisal:2023fgb, Bisal:2023iip}).

\item $\mathtt{MicrOMEGAs}$ can calculate the SI-DD $\tilde{\chi}_1^0-N$ cross-sections at the LO through the $\tilde{\chi}_1^0-q$ and $\tilde{\chi}_1^0-g$ scattering process. 
    The $\tilde{\chi}_1^0-q$ process is mediated by the tree-level $\tilde{\chi}_1^0\tilde{\chi}_1^0h_i$ vertex whereas the $\tilde{\chi}_1^0-g$ scattering takes place through the one-loop diagrams shown in Fig.~\ref{fig:boxdiag_sq}. In $\mathtt{MicrOMEGAs}$, the LO $\tilde{\chi}_1^0-g$ scattering is included in terms of effective interactions in the low-energy limit, following the Ref.~\cite{Drees:1992rr, Drees:1993bu}. We denote this LO cross-section by $\sigma_{\rm SI}^{\rm LO}$.
    
    \item We modify the $\tilde{\chi}_1^0\tilde{\chi}_1^0h_i$ vertex within the $\mathtt{MicrOMEGAs}$ by incorporating all the triangular topologies shown in~Fig.~\ref{fig:topology1}. 
    
    \item Additionally, we include the one-loop and two-loop diagrams shown in Fig.~\ref{fig:boxdiag}b and \ref{fig:twoloop} to evaluate the quark \textbf{twist-2} and gluon contributions to $\lambda_G$, respectively. Their analytical expressions are available in the literature but were not adapted in $\mathtt{MicrOMEGAs}$. In this study, we use the analytical expressions for the two-loop gluon (Fig.~\ref{fig:twoloop}b and \ref{fig:twoloop}c) and quark \textbf{twist-2} contributions (Fig.~\ref{fig:boxdiag}b) from Ref.~\cite{Hisano:2010fy, Hisano:2010ct}.
    Note that in the two-loop diagram in Fig.~\ref{fig:twoloop}a, only $W$ boson loop has been considered in the literature. 
    Now that we include all the MSSM particles contributing to the $\tilde{\chi}_1^0\tilde{\chi}_1^0h_i$ vertex, the modification in diagram Fig.~\ref{fig:twoloop} occurs accordingly. 

  In this improved analysis, we invoke all the contributions corresponding to the diagrams shown in Fig.~\ref{fig:boxdiag}, \ref{fig:twoloop}, and \ref{fig:topology1}, along with the existing LO contributions. Subsequently, the total cross-section of scattering may be 
   referred to as the NLO cross-section, denoted by $\sigma_{\rm SI}^{\rm NLO}$.

\end{itemize}

\textbf {Validation of the amplitudes at NLO\,:} Before depicting the NLO-improved DM-nucleon SI cross-section, it is customary to learn the individual amplitudes of $\tilde{\chi}_1^0$-nucleon scattering through the NLO corrections of $\tilde{\chi}_1^0\tilde{\chi}_1^0h_i$ vertex, gluon NLO, and quark \textbf{twist-2} contributions. The overall SI cross-section at NLO is simply obtained through their individual strength and interference effects.

For the sake of validation, apart from the LO amplitude,
we classify the complete loop contributions through (i)
Higgs and squark mediated amplitudes (see Fig.~\ref{fig:boxdiag}a, \ref{fig:boxdiag_sq}, \ref{fig:twoloop}a and \ref{fig:topology1}) (ii) quark \textbf{twist-2} (see Fig.~\ref{fig:boxdiag}b) and (iii) gluon two-loop scalar contributions (see Fig.~\ref{fig:twoloop}b,~\ref{fig:twoloop}c). Neglecting the squark contributions 
for heavy squark masses, while
depicting the numerical checks, we further assemble (i)+(iii) along with the tree-level DM-nucleon scattering as \textbf{vertex\,+\,gluon} contributions, which are 
determined by the scalar couplings $\lambda_q$ and $\lambda_G$ up to NLO. 

\begin{figure}[H]
	\centering
     \hspace{-1.0cm}\includegraphics[width=0.54\linewidth]{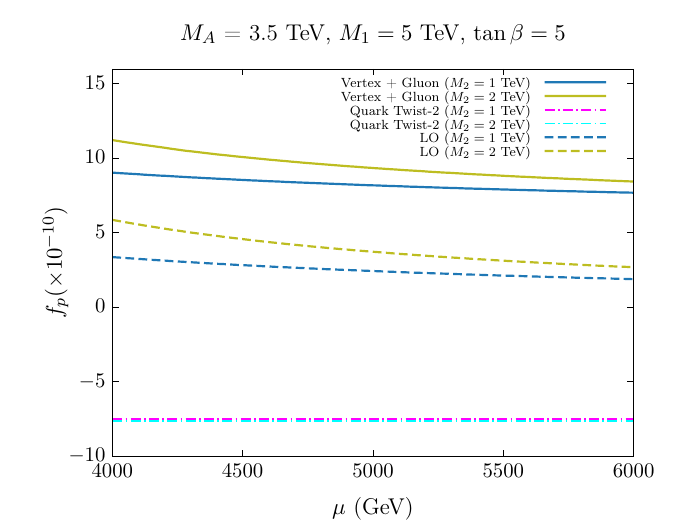}	
     \hspace{-0.4cm}\includegraphics[width=0.54\linewidth]{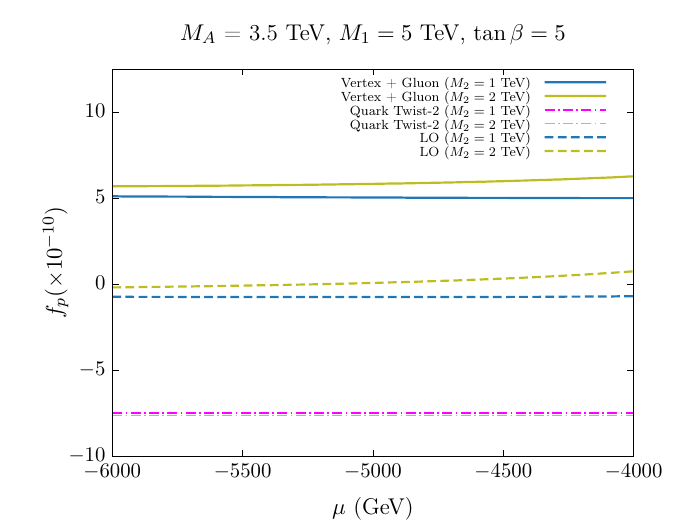}\\
     (a)\hspace{8.0cm}(b)\\
     \includegraphics[width=0.54\linewidth]{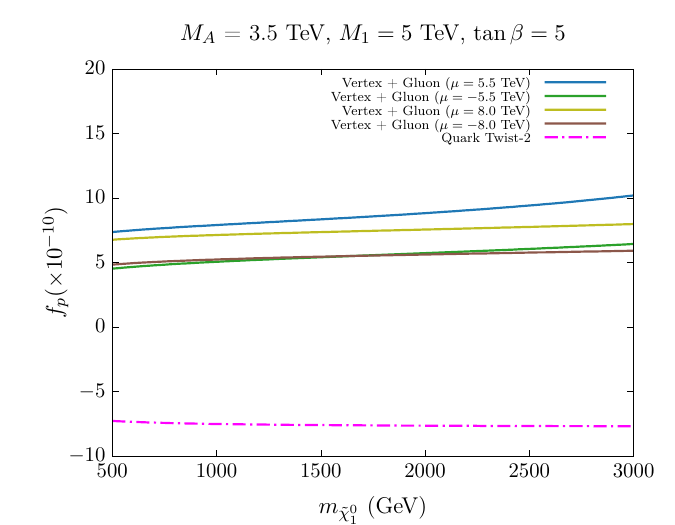}\\
     \qquad\,\,\,\,(c)
	\caption{Figures (a) and (b) show the variation of the $\tilde{\chi}_1^0$-nucleon (proton) amplitude with the $\mu$ parameter, originated from different LO and NLO contributions in Fig.~\ref{fig:boxdiag}-\ref{fig:topology1} (for details, see the text). Here, $M_2=1$~TeV and $2$~TeV are assumed for both the cases $\mu>0$ and $\mu<0$ while $\tan\beta=5$ is fixed. Similarly, we set the other parameters $M_A=3.5$~TeV, $m_{\tilde{\ell}_{L, R}}=3.5$~TeV, and $m_{\tilde{q}_{L, R}}=7$~TeV. Similarly, figure (c) shows the variation of the same with $m_{\tilde{\chi}_1^0}$ where $\mu$ is fixed at $\pm 5.5$~TeV and $\pm 8.0$~TeV, respectively. Note that the quark \textbf{twist-2} and gluon two-loop contributions show negligible dependence on $M_2$ or $|\mu|$. In figure (c), we display a single line for the quark \textbf{twist-2} contributions because the contributions overlap for the chosen $\mu$ values, making them visually indistinguishable.}
	\label{fig:ampl}
\end{figure}

We begin discussion with Fig.~\ref{fig:ampl}a and \ref{fig:ampl}b, that illustrate the variation of the $\tilde{\chi}_1^0$-nucleon (proton) scattering amplitude, $f_p$ with $\mu$ for the combinations alongside LO, stated previously, where $M_2$ is fixed at $1$~TeV and $2$~TeV, respectively. 
Fig.~\ref{fig:ampl}c shows the variations of the same with $m_{\tilde{\chi}_1^0}$, assuming $\mu = \pm 5.5$~TeV and $\pm 8.0$~TeV. For all the plots, $\tan\beta=5$ is taken.
In Fig.~\ref{fig:ampl}a and \ref{fig:ampl}b, the \textbf{vertex\,+\,gluon} contribution to the amplitude, $f_p$, decreases gradually as $|\mu|$ increases. This is because as $\mu$ increases, the Wino-Higgsino mixing decreases for a fixed value of $M_2$. Consequently, the LO contribution to $\tilde{\chi}_1^0\tilde{\chi}_1^0h$ vertex decreases, resulting in a reduction of the vertex contributions. Another interesting observation is that for $\mu<0$, the \textbf{vertex\,+\,gluon} amplitude is relatively smaller than for $\mu>0$, as is evident from all the figures in \ref{fig:ampl}a, \ref{fig:ampl}b, and \ref{fig:ampl}c. This is because, for $\mu<0$, the LO $\tilde{\chi}_1^0\tilde{\chi}_1^0h$ becomes much smaller (see the discussion below), which reduces the vertex contributions. Note that in all the scenarios, the quark \textbf{twist-2} and gluon two-loop contributions have negligible dependence on $M_2$ or $|\mu|$ as we consider the pure Wino limit for their amplitudes. The Wino fraction is $>99.9\%$ in all the cases, thus consistent with our assumption. 
At this point, we can compare our results with the earlier work in Ref.~\cite{Hisano:2010fy}. In Fig.~4 of that reference, the individual contributions to the Wino-nucleon amplitude are plotted. For instance, for $m_{\tilde{\chi}_1^0}=1.1$~TeV, the \textbf{vertex\,+\,gluon} contribution (or Higgs $+$ two-loop, as per their terminology) can be approximately anticipated to be $6.5 \times 10^{-10}$ in the pure Wino limit for $m_h =125~$GeV. According to our calculations, from Fig.~\ref{fig:ampl}c,  this value becomes $7.2 \times 10^{-10}$, resulting in a $\sim 10\%$ enhancement in the \textbf{vertex\,+\,gluon} contribution. Here we assume a higher value of 
$\mu=8~$TeV for comparison. Since we are using the same contributions for the DM-gluon scattering as in Ref.~\cite{Hisano:2010fy}, this $\sim 10\%$ enhancement can be attributed to the additional but complete set of one-loop EW corrections (see Fig.~\ref{fig:topology1}), which have not been considered earlier.

\textbf{Observations and outcomes of NLO corrections\,:} In the overall amplitude, one can observe a cancellation between the \textbf{vertex\,+\,gluon} and the quark \textbf{twist-2} contributions (see also Ref.~\cite{Hisano:2010fy, Hisano:2011cs, Hisano:2012wm}). The feature holds for any $\mu$ values though the degree of cancellation differs. In the following, we discuss both cases, i.e., $\mu>0$ and $\mu<0$ scenarios.
\begin{itemize}
\item \underline{$\mu>0$ scenario\,:} Following Fig.~\ref{fig:ampl}a and Fig.~\ref{fig:ampl}c, we observe a strong cancellation between the \textbf{vertex\,+\,gluon} and \textbf{twist-2} contributions. The cancellation can lead to unfavorable occasions where the NLO cross-section becomes lower than the LO value or even produces vanishingly small contributions. The latter is usually referred to as \textbf{blind spot} scenario in the literature~\cite{Cheung:2012qy}~\footnote{ Another example follows when the SM-like Yukawa couplings of the light quarks are relaxed~\cite{Das:2020ozo}.}. We will substantiate this in the following section. In Fig.~\ref{fig:ampl}c, we find that the cancellation is most conspicuous at $\mu = 5.5$~TeV when $m_{\tilde{\chi}_1^0} \simeq 500$~GeV, which refers to the Higgsino fraction of $\sim 0.02\%$.
Similarly, for $\mu=8$~TeV, the cancellation leaves the maximum impact around $m_{\tilde{\chi}_1^0}\simeq 2.25$~TeV where the Higgsino fraction is $\sim 0.02\%$. 
Due to the cancellation, the $\tilde{\chi}_1^0$-nucleon amplitude at NLO, and consequently the corresponding SI cross-section, will be highly suppressed and may reach the \textbf{blind spots} at NLO.  
For larger values of $\mu$ (e.g., $\mu = 9.50$~TeV), the maximum cancellation in the DM-nucleon NLO amplitude or \textbf{blind spots} will shift to higher values of $m_{\tilde{\chi}_1^0}$ (e.g., $m_{\tilde{\chi}_1^0} \simeq 3.65$~TeV) while the Higgsino fraction lives in the same ballpark as before. Thus, typically, we may say that the \textbf{blind spots} or maximum degree of cancellations favor a tiny Higgsino component of the order of $\sim \mathcal{O}(0.02\%)$ within the Wino-like LSP.
Overall, we observe that the SI-DD NLO amplitude will hardly receive any significant boost compared to the LO results, rather it supports the null results that we are experiencing in the different DM
DD experiments so far.

  \item \underline{$\mu<0$ scenario\,:}  
Here, the leading-order $\tilde{\chi}_1^0$-nucleon amplitude is quite small, which primarily follows from Eq.~\eqref{clcrlo1}. There we see that $C_{L,R}^{\rm LO}(h)\propto \big[M_2 + \mu \sin\big(2\beta\big)\big]$. 
Thus, for $\mu<0$, the LO coupling and, consequently, the amplitude reaches its minimum when
    \begin{align}
   &M_2 + \mu \sin\big(2\beta\big)=0 \hskip1.5cm \big[{\rm For}\,\, M_2\neq |\mu|\big]\\
    \implies & \mu = - \frac{M_2\big(1+\tan^2\beta\big)}{2\tan\beta}~.
    \label{minlocoupling}
\end{align}
For instance, if we set $M_2 = 1$~TeV (leading to $m_{\chi_1^0} \sim 1$~TeV) and $\tan\beta = 5$, the LO amplitude reaches its minimum or produces \textbf{blind spots} at $\mu = -2.6$~TeV. Similarly, for $M_2 = 2.5$~TeV and $\tan\beta = 5$, the \textbf{blind spots} are observed at $\mu = -6$~TeV. Overall, for negative $\mu$ values, the cancellation in $C^{\rm LO}_{L, R}$ causes the LO DM-nucleon amplitude to remain lower (can be also negative, see e.g., Fig.~\ref{fig:ampl}b) compared to the positive $\mu$ values of the same magnitude, and, in the \textbf{blind spot} regions, one finds that the LO DM-nucleon amplitude becomes vanishingly small. This may
clearly be visible from Fig.~\ref{fig:ampl}a and \ref{fig:ampl}b. Upon including the full NLO corrections, the \textbf{vertex\,+\,gluon} contributions may dilute or boost the DM-nucleon cross-section after cancellation with the quark \textbf{twist-2} operators in different parts of the MSSM parameter space.   
Specifically, in the regime where the LO cross-section approaches zero for $\mu < 0$, the NLO cross-section effectively mitigates the effects. Similarly, with non-zero and positive
$C^{\rm LO}_{L,R}$, the NLO cross-section may be reduced compared to the LO value, due to destructive interference between the quark \textbf{twist-2} and \textbf{vertex\,+\,gluon} contributions.
We further confirm this in the next section (Sec.~\ref{sec:numericalres}).
\end{itemize}

\section{Numerical Results}
\label{sec:numericalres}
In this section, we delineate the numerical results that illustrate the influence of the DM DD cross-section on the MSSM parameter space, which arises from the one-loop corrections to the Wino-like $\tilde{\chi}_1^0 \tilde{\chi}_1^0 h$ and $\tilde{\chi}_1^0 \tilde{\chi}_1^0 H$ vertices as well as one-loop $\tilde{\chi}_1^0\tilde{\chi}_1^0\bar{q}q$, $\tilde{\chi}_1^0\tilde{\chi}_1^0gg$ and two-loop $\tilde{\chi}_1^0\tilde{\chi}_1^0gg$ processes.
For assessing the dependence of $C^{\rm NLO}_{L,R}$ (see Eq.~\eqref{eqn:nlo}) and other contributions 
(Fig.~\ref{fig:boxdiag}-\ref{fig:twoloop})
 numerically, we perform a comprehensive scan for $\tan\beta=5$, over the following ranges of the parameters\,:
\begin{align}
& 0.5\,\,{\rm TeV} \le M_2 \le 3\,\,{\rm TeV},\nonumber\\
& 0.5\,\,{\rm TeV} \le M_1 \le 15\,\,{\rm TeV},\nonumber\\
& 0.5\,\,{\rm TeV} \le \lvert\mu\rvert \le 15\,\,{\rm TeV}.
\label{pararanges}
\end{align}

We vary these parameters randomly over the specified ranges, imposing the condition, $M_2<M_1, |\mu|$. 
We set the other parameters $M_A=3.5$~TeV and $m_{\tilde{\ell}_{L, R}}=3.5$~TeV  while permitting both positive and negative values for the Higgsino mass parameter $\mu$. Also, we set $m_{\tilde{q}_{L, R}}=7$~TeV and $\tan\beta=5$.
The high value of
the squark masses helps to validate the parameter
space against the LHC constraints. The large
masses of the sleptons are assumed to allow the
LSP, completely determined by $\tilde \chi_1^0$,
even for a relatively larger value of the LSP mass (say, for instance, up to 3 TeV).
We allow all the points to satisfy the $B$-physics constraints at $2\sigma$ variations, e.g., $ 3.02 \times 10^{-4} < BR(b \to s\gamma) < 3.62 \times 10^{-4}$~\cite{HFLAV:2019otj}, 
$2.23 \times 10^{-9} < BR(B_s\to \mu^+\mu^-) <  3.63 \times 10^{-9}$~\cite{Altmannshofer:2021qrr}.~\footnote{As an aside, we consider the on-shell renormalization scheme to calculate the one-loop mass splitting between $\tilde{\chi}_1^\pm$ and $\tilde{\chi}_1^0$ \big[see Eq.~\eqref{eq:onshellsplitting}\big], which comes out to be $\Delta m(\tilde{\chi}_1^\pm, \tilde{\chi}_1^0)\simeq 160$~MeV corresponding $\tau_{\tilde{\chi}_1^\pm} = 0.2 \rm ns$.} In calculating the SUSY Higgs mass, we account for a 3 GeV theoretical uncertainty, which gives rise to the following range for the SM-like Higgs mass in the MSSM~\cite{Bahl:2019hmm, ATLAS:2015yey, ATLAS:2012yve, CMS:2012qbp}:
\begin{align}
    122~{\rm GeV}<m_h<128~{\rm GeV}.
\end{align}

From this
parametric scan, we only focus on the DM, which is
predominantly Wino-like while minimal to moderate
Higgsino components may be present. In the first place, we refrain from imposing constraints on the relic density. 
Here, we are mainly interested in the (i) enhancement in the coupling $\tilde{\chi}_1^0\tilde{\chi}_1^0h_i$ relative to the tree-level values, and (ii) improved SI-DD cross-sections induced by the one-loop radiative corrections to the corresponding vertex as well as other one-loop and two-loop contributions via Fig.~\ref{fig:boxdiag}-\ref{fig:twoloop}, as discussed earlier. 
However, later, we tabulate a few benchmark points (BMPs) in Tab.~\ref{tab:bmplabel} to take note
of our important results, where all the existing
theoretical and experimental checks would be satisfied. 
For experimental inputs, we use the results from \textbf{LUX-ZEPLIN (LZ)} (2022), \textbf{XENON1T} (2018), and \textbf{LUX} (2017), among which \textbf{LZ} provides the most stringent constraints for the SI-DD cross-sections.

\begin{figure}[H]
	\centering
	\includegraphics[width=0.46\linewidth]{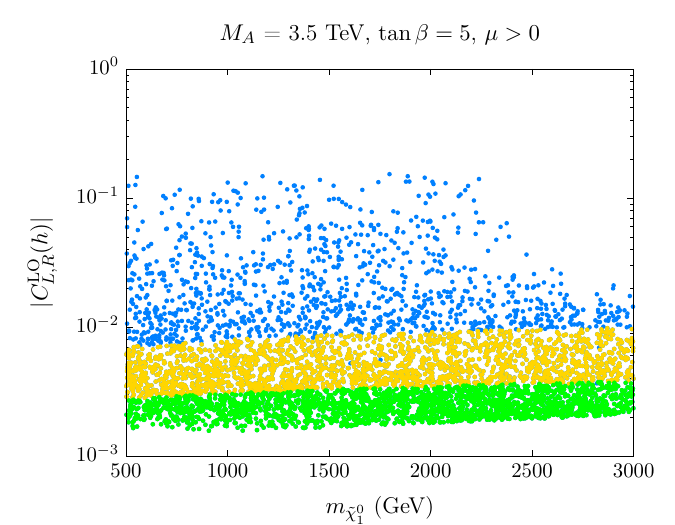}	
    \includegraphics[width=0.46\linewidth]{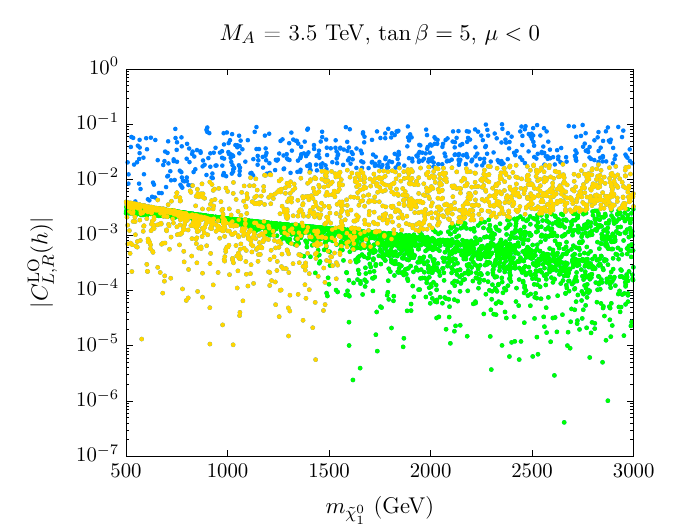}\\
     \quad\quad\quad(a)\hspace{6.3cm}\quad\,\,\quad\,\,(b)
	\caption{Figures (a) and (b) depict the variations of the LO $\tilde{\chi}_1^0\tilde{\chi}_1^0h$ coupling with the LSP mass for $\mu>0$ and $\mu<0$, respectively. We set $\tan\beta=5$, $M_A=3.5$~TeV, $m_{\tilde{\ell}_{L, R}}=3.5$~TeV, and $m_{\tilde{q}_{L, R}}=7$~TeV. The other input parameters, which are varied, are mentioned in the text. In figure (a), the green region corresponds to Wino component with $|\mathcal{N}_{12}|^2\geq 99.99\%$, the yellow region to $99.95\%\leq |\mathcal{N}_{12}|^2\leq 99.99\%$, and the blue region to $90\%\leq |\mathcal{N}_{12}|^2\leq 99.95\%$. In figure (b), the same color palette zones represent Wino components of $|\mathcal{N}_{12}|^2\geq 99.95\%$, $99.50\%\leq |\mathcal{N}_{12}|^2\leq 99.95\%$, and $90\% \leq |\mathcal{N}_{12}|^2\leq 99.50\%$, respectively. All the points satisfy the $B$-physics and Higgs mass constraints. }
	\label{fig:LOh10401}
\end{figure}

In the presentation of the numerical results, we begin with demonstrating the parametric dependence of the LO $\tilde{\chi}_1^0\tilde{\chi}_1^0h$ coupling, $C^{\rm LO}_{L, R}(h)$, on the LSP mass.
Fig.~\ref{fig:LOh10401}a and \ref{fig:LOh10401}b depict the variations of the LO couplings for $\mu>0$ and $\mu<0$, respectively for a fixed $\tan\beta=5$ and $M_A=3.5~$TeV.
In Fig.~\ref{fig:LOh10401}a, the green region has the maximum Wino fraction, with $|\mathcal{N}_{12}|^2\geq 99.99\%$, the yellow region has $99.95\%\leq |\mathcal{N}_{12}|^2\leq 99.99\%$, and the blue region has $90\%\leq |\mathcal{N}_{12}|^2\leq 99.95\%$. 
This qualitative distinction allows $C^{\rm LO}_{L, R}$ to reach up to a value $\sim 10^{-3}$, $\sim 10^{-2}$, $\sim 10^{-1}$ for the green, yellow, and blue regions, respectively.
In Fig.~\ref{fig:LOh10401}b, 
$C_{L,R}^{\rm LO}$ can maximally reach the similar values as in Fig.~\ref{fig:LOh10401}a through the same color-classified zones. 
Here, the green region has $|\mathcal{N}_{12}|^2\geq 99.95\%$, the yellow region has $99.50\%\leq |\mathcal{N}_{12}|^2\leq 99.95\%$, and the blue region has $90\% \leq |\mathcal{N}_{12}|^2\leq 99.50\%$. 
However, for the lowest reach, the $\tilde{\chi}_1^0\tilde{\chi}_1^0h$ coupling for $\mu<0$ may become vanishingly small. 
This follows from the fact that for $\mu>0$, $C^{\rm LO}_{L,R}(h)$
only takes the lower values when $|\mu| >> M_2$, i.e., especially for higher Wino fraction when the LSP approaches a pure Wino-like state. 
In contrast, for $\mu<0$, a complementary contribution arises from the relative sign difference between $M_2$ and $\mu$.  
Together, these two effects result in a greater reduction of $C_{L, R}^{\rm LO}$ for $\mu<0$ compared to $\mu>0$ despite the absolute value of the Higgsino mass parameter being the same. 
The same reason may lead to \textbf{blind spots} in the first case, as discussed earlier.

\begin{figure}[H]
	\centering
   \includegraphics[width=0.468\linewidth]{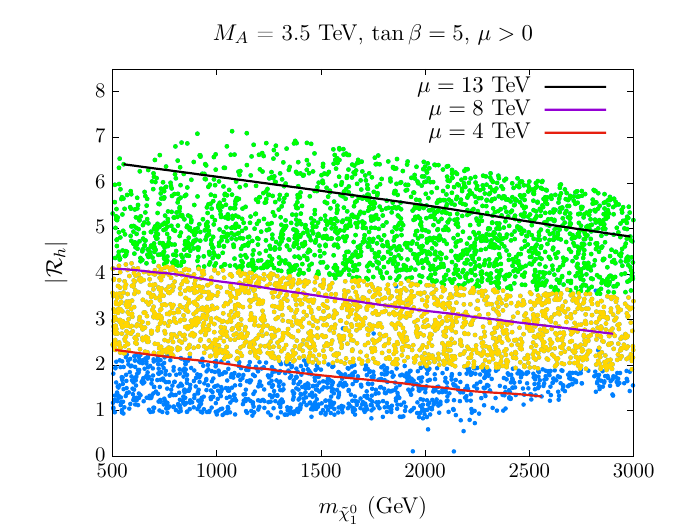}
   \includegraphics[width=0.468\linewidth]{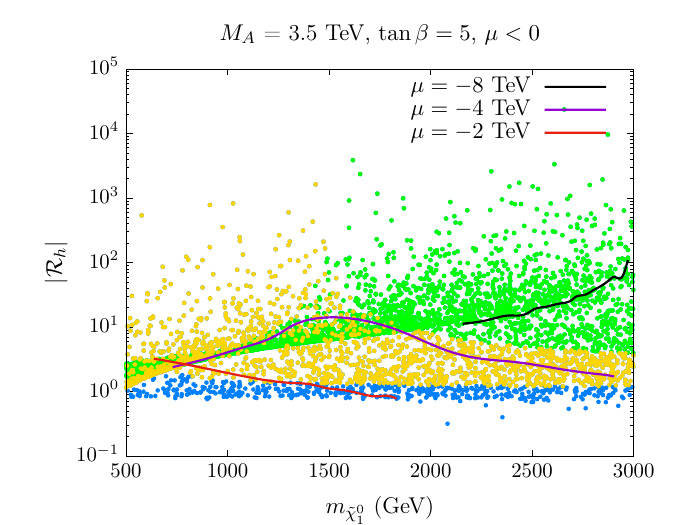}\\
    \qquad\quad(a)\hspace{7.3cm}(b)\\
  \caption{Figures (a) and (b) show the variations of $|\mathcal{R}_h|$ with $m_{\tilde{\chi}_1^0}$ for $\tan\beta = 5$, in the cases of $\mu>0$ and $\mu<0$, respectively. All the other input parameters are the same as in the previous case. The color codes are the same as in Fig.~\ref{fig:LOh10401}a and \ref{fig:LOh10401}b for $\mu>0$ and $\mu<0$, respectively. All the points satisfy the $B$-physics and Higgs mass constraints.}
	\label{fig:hclHclsigmaSIp}
\end{figure}

\begin{figure}[H]
	\centering
    \includegraphics[width=0.468\linewidth]{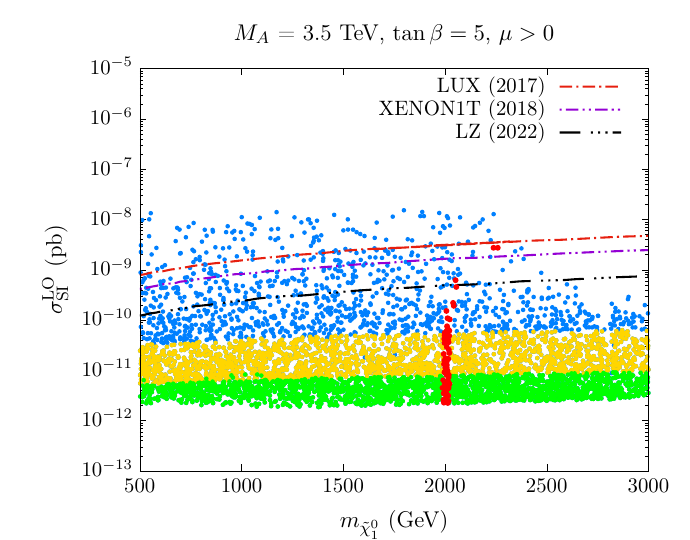}
    \includegraphics[width=0.468\linewidth]{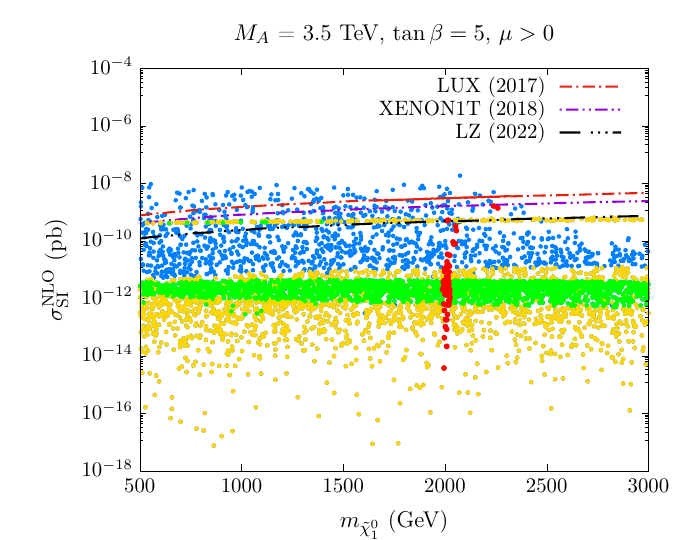}\\
      \,\qquad\quad(a)\hspace{7.3cm}(b)\\
  \includegraphics[width=0.468\linewidth]{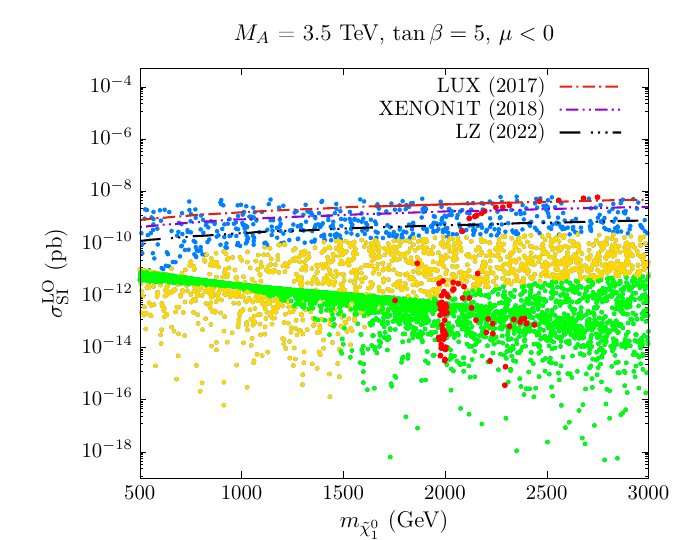}
	\includegraphics[width=0.468\linewidth]{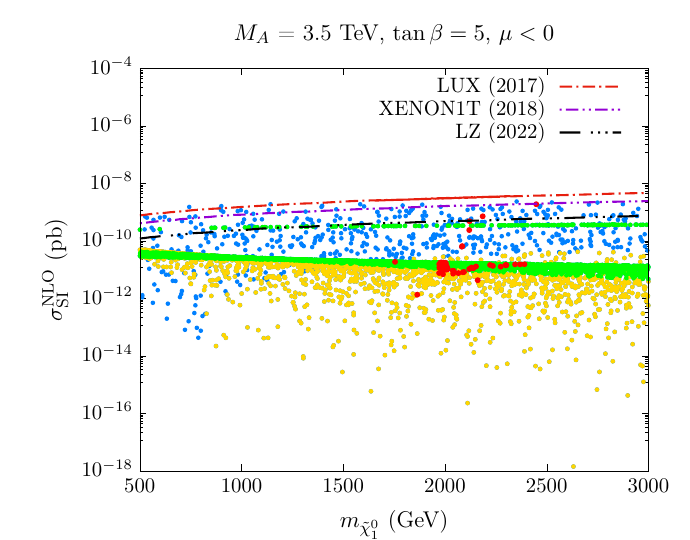}\\
   \quad\qquad(c)\hspace{7.2cm}(d)
  \caption{Figures (a) and (b) represent the variations of $\sigma_{\rm SI}^{\rm LO}$ and $\sigma_{\rm SI}^{\rm NLO}$ with $m_{\tilde{\chi}_1^0}$, for $\tan\beta=5$, $\mu>0$. Figures (c) and (d) depict the same features but for $\mu<0$.
   All the other input parameters are the same as in the previous case. The color codes are the same as in Fig.~\ref{fig:LOh10401}a and \ref{fig:LOh10401}b for $\mu>0$ and $\mu<0$, respectively.
  The red points satisfy the relic density within the $2\sigma$ variation. Note that in the computation of $\sigma_{\rm SI}^{\rm NLO}$, the relic density has been calculated using the NLO-corrected LSP-Higgs vertices, whereas in the computation of $\sigma_{\rm SI}^{\rm LO}$, the tree-level LSP-Higgs vertices have been used. All the points satisfy the $B$-physics and Higgs mass constraints.}
	\label{fig:hclHclsigmaSIpc}
\end{figure}

The LO $\tilde{\chi}_1^0\tilde{\chi}_1^0h$ coupling receives the complete set of one-loop EW corrections and the counterterm contributions through Eq.~\eqref{eqn:nlo}.
It is noteworthy that when the LSP is primarily Wino-like with a negligible Higgsino component (for instance, the green regions in Fig.~\ref{fig:LOh10401}), apart from Fig.~\ref{fig:topology1}b, the contributions from the remaining one-loop vertices in Fig.~\ref{fig:topology1} usually have minimal impacts on $C^{\rm NLO}_{L, R}$ for our choice of the parameters. 
However, when the LSP has a non-negligible Higgsino component, the vertices involving virtual Higgs bosons, EW fermions, and the corresponding counterterms may contribute significantly towards NLO improved vertices. 
To provide a clearer illustration, we can define the enhancement in the $\tilde{\chi}_1^0\tilde{\chi}_1^0h$ coupling as $|\mathcal{R}_h|=\Big|\frac{C^{\rm NLO}_{L, R}}{C^{\rm LO}_{L, R}}\Big|$. 
We depict the variation of $|\mathcal{R}_h|$ with $m_{\tilde\chi_1^0}$ in Fig.~\ref{fig:hclHclsigmaSIp}.
The green, yellow, and blue regions in Fig.~\ref{fig:hclHclsigmaSIp}a and \ref{fig:hclHclsigmaSIp}b correspond to the same Wino fractions as we have in Fig.~\ref{fig:LOh10401}a and 
Fig.~\ref{fig:LOh10401}b respectively.
We also show the contours for the Higgsino mass parameter. The black, brown, and red contours correspond to $\mu=13~(-8)$~TeV, $8~(-4)$~TeV, and $4~(-2)$~TeV, in Fig.~\ref{fig:hclHclsigmaSIp}a (Fig.~\ref{fig:hclHclsigmaSIp}b), respectively. Following the discussion in Sec.~\ref{sec:improvementMicro} (Fig.~\ref {fig:ampl}), we observe that $\mu \in[5.5,8]$ TeV is desired to have a maximum drop in the NLO cross-section. In Fig.~\ref{fig:hclHclsigmaSIp}, we see that the yellow region falls in that part of the parameter space where, for $\mu>0$, it is characterized by $|\mathcal R_h|\in [2,\,4]$. Additionally, we note that for $\mu >0$, $\mathcal R_h >0$ always. Thus the NLO vertex corrections 
$C^{\rm NLO}_{L, R}$ boosts the \textbf{vertex\,+\,gluon} part in the NLO cross-section, which in turn leads to the reduction of overall NLO cross-sections. 
From the blue to green region, the Wino fraction tends to increase, resulting in a lower value for $C^{\rm LO}_{L, R}$, which 
in turn leads to a higher value for $|\mathcal{R}_h|$. 
As discussed earlier, $\mu<0$ lowers the LO $\tilde{\chi}_1^0\tilde{\chi}_1^0h$ coupling significantly throughout the parameter space, resulting in a quite high rise in $|\mathcal{R}_h|$. Finally, the resultant cross-sections are shown in Fig.~\ref{fig:hclHclsigmaSIpc}.

Fig.~\ref{fig:hclHclsigmaSIpc}a and \ref{fig:hclHclsigmaSIpc}b demonstrate the variations of the LO and NLO cross-sections with $m_{\tilde{\chi}_1^0}$ for $\tan\beta=5$ and $\mu>0$, while Fig.~\ref{fig:hclHclsigmaSIpc}c and \ref{fig:hclHclsigmaSIpc}d show the same for $\mu<0$.
We begin with the $\mu>0$ scenario. 
As expected, the LO cross-sections become much smaller when the DM becomes more Wino-like. For instance, the green or yellow region in Fig.~\ref{fig:hclHclsigmaSIpc}a has $\sigma_{\rm SI}^{\rm LO}\sim \mathcal{O}\big(10^{-12}-10^{-11}\big)$~pb, which resides well below the \textbf{LZ} line. 
For the blue points, the Wino-Higgsino mixing is most prominent, indicating a substantial contribution from the tree-level $\tilde{\chi}_1^0\tilde{\chi}_1^0h$ vertex. 
The LO cross-section is most promising for this region.
The NLO estimates, as shown in Fig.~\ref{fig:hclHclsigmaSIpc}b, indicate a lower cross-section for most of the points in general, though it is prominently visible for the yellow points. 
Now, as discussed in Sec.~\ref{sec:improvementMicro}, the overall fall in NLO cross-sections can be comprehended from the fact that there is a cancellation between the quark \textbf{twist-2} contributions and all other contributions (see Fig.~\ref{fig:ampl}). 
As shown, a minimal Higgsino component 
with the Higgsino mass parameter $\mu \in [5.5,8.0]\text{~TeV}$ particularly favors the conspiracy in the maximum cancellation among different contributory terms, which mostly coincides with the yellow region. 
Consequently, the NLO cross-sections for this region become small, nearly vanishing for some points in the parameter space across the entire range of Wino masses we have considered. 
Therefore, we observe cross-sections approaching zero or \textbf{blind spots} at NLO for some of the points. 
The \textbf{blind spots} can also be approached for $\mu<0$ scenario, but mainly at the LO, as shown in Fig.~\ref{fig:hclHclsigmaSIpc}c (see the discussion in Sec.~\ref{sec:improvementMicro}). 
Here, we observe that the LO cross-sections for Wino-like DM, represented by the green region, along with some parts of the yellow region, reach the lowest values. After incorporating the NLO EW corrections, the green region has been significantly uplifted, and the \textbf{blind spots} for this region have disappeared. 
The results can be seen in Fig.~\ref{fig:hclHclsigmaSIpc}d.
In contrast, a few points in the yellow region still exhibit low NLO cross-section, indicating the presence of the cancellation, albeit by a lesser amount, even after the inclusion of NLO EW corrections. 
As an aside, in Fig.~\ref{fig:hclHclsigmaSIpc}, all the points satisfy the $B$-physics and Higgs mass constraints. 
The red points satisfy the relic density within the $2\sigma$ variation of the central value $\Omega h^2=0.1198$.
The relic density is satisfied for $m_{\tilde{\chi}_1^0}\sim 2$~TeV. Note that in the computation of $\sigma_{\rm SI}^{\rm NLO}$, the relic density has been calculated using the NLO-corrected LSP-Higgs vertices, whereas in the computation of $\sigma_{\rm SI}^{\rm LO}$, the tree-level LSP-Higgs vertices have been used.  
We have also checked that a deviation of up to $10\%$ may be found in the relic density after incorporating the NLO corrections.
For earlier studies on radiative corrections to the relic density, see, e.g., Ref.~\cite{Baro:2007em, Baro:2009na, Chatterjee:2012hkk, Ciafaloni:2013hya, Beneke:2014gla, Beneke:2016ync, Harz:2014tma, Herrmann:2014kma, Boudjema:2014gza}.

After highlighting the details of the DM-nucleon SI cross-section at the NLO, for a relatively low value of 
$\tan\beta$, we have also considered a relatively high $\tan\beta(=30)$. 
In particular, it mainly affects in two places\,: (i) the LO $C^{\rm LO}_{L, R}$ coupling, where varying $\tan\beta$ for a given $M_2$ shifts the value of the $\mu$ parameter necessary to achieve LO \textbf{blind spots} for $\mu<0$, and (ii) the DM DD cross-section, which grows with $\tan\beta$. 
However, if we restrict the Higgsino fraction to $\leq 10\%$ (as for $\tan\beta=5$), our discussions and observations remain unchanged, and thus, we do not present this for further discussion.\\

\begin{figure}[H]
	\centering
	\includegraphics[width=0.62\linewidth]{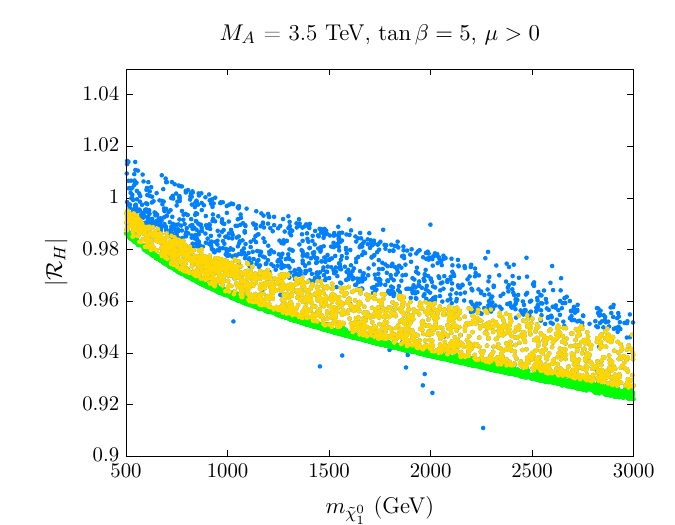}
	\caption{Variations of $|\mathcal{R}_H|$ with $m_{\tilde{\chi}_1^0}$ for $\mu>0$. The color codes are the same as in Fig.~\ref{fig:LOh10401}a. All the points satisfy the $B$-physics and Higgs mass constraints.}
	\label{fig:couplinghclne}
\end{figure}

Since we have incorporated the contributions to the DM DD cross-section arising from the CP-even heavy Higgs-mediated process, it is pertinent to highlight a few key points regarding this. Importantly, the contributions to the SI-DD cross-section from the heavier CP-even Higgs-mediated process are substantially lower than those from the SM-like Higgs for our choice of the parameters.
In Fig.~\ref{fig:couplinghclne}, we observe that for most points, $|\mathcal{R}_H| < 1$ (where  $|\mathcal{R}_H|=\Big\lvert\frac{C_{L,R}^{\rm NLO}}{C_{L,R}^{\rm LO}}\Big\rvert_H $ is the ratio of NLO and LO couplings for the $\tilde{\chi}_1^0\tilde{\chi}_1^0H$ vertex), except for a few blue points in the region where $m_{\tilde{\chi}_1^0} < 1$~TeV.  
This means that the LO $\tilde{\chi}_1^0 \tilde{\chi}_1^0 H$ vertex interferes destructively with its one-loop corrections, leading to a NLO corrected coupling that is smaller than the LO value. 
A similar behavior is observed in the $\mu < 0$ scenario (not shown), where $|\mathcal{R}_H| < 1$ for all points, indicating that every point undergoes negative corrections. 
We note that when the heavy Higgs mass is at the multi-TeV scale, its contribution to the SI cross-section through the heavy Higgs-mediated $\tilde{\chi}_1^0\tilde{\chi}_1^0 \bar{q}q$ process becomes negligible. This follows from the fact that SI cross-section scales as $\sigma_{\rm SI}\propto \mathcal{O}\left(\frac{1}{m_{h,H}^4}\right)$.

\begin{figure}[H]
	\centering
	\includegraphics[width=0.58\linewidth]{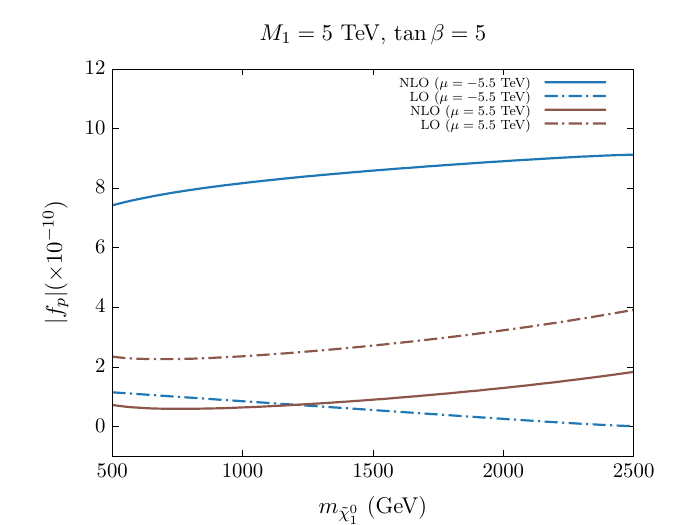}
	\caption{Variations of the NLO and LO amplitudes for $\tilde{\chi}_1^0$-nucleon scattering with $m_{\tilde{\chi}_1^0}$, considering relatively lower values of the heavy Higgs mass, $m_H \in [450, 1200]$~GeV.}
	\label{fig:ampllh}
\end{figure}

If the heavy Higgs masses are assumed to have relatively low values, the contributions from processes mediated by the heavy Higgs can become significant. For instance, Fig.~\ref{fig:ampllh} demonstrates the variation in the $\tilde{\chi}_1^0$-nucleon scattering amplitude assuming the heavy Higgs masses in the range, $m_H \in [450, 1200]~\text{GeV}$. For $\mu < 0$, we observe \textbf{blind spots} in the LO calculations around $m_{\tilde{\chi}_1^0} \sim 2.5~\text{TeV}$, where the amplitudes (and consequently the cross-sections) vanish.
We have checked that this vanishing occurs due to destructive interference between the tree-level processes mediated by the SM-like Higgs and the heavy Higgs. This behavior can be understood from Eq.~\eqref{clcrlo1} and \eqref{clcrHlo2}, where the LO $\tilde{\chi}_1^0\tilde{\chi}_1^0h$ and $\tilde{\chi}_1^0\tilde{\chi}_1^0H$ couplings receive opposite signs for some specific values of the parameters. For instance, for $\tan\beta=5$ and $\mu<0$, the couplings have opposite signs when $M_2+ \mu\sin (2\beta) >0$. Importantly, after incorporating NLO corrections, the \textbf{blind spots} disappear.
For $\mu > 0$, there are no \textbf{blind spots} at LO. In this case, the contributions from the SM-like and heavy Higgs bosons interfere constructively. However, after incorporating NLO corrections, the parameter space points approach \textbf{blind spots} for $m_{\tilde{\chi}_1^0} \sim 700$~GeV under the chosen parameters.

\begin{table}[H]
	\centering
		\begin{tabular}{|c|c|c|c|c|c|c|c|c|c|c|c|c|}

			\hline
			\multicolumn{8}{|c|}{$\mu>0$ scenario} \\ 		
			\hline
			
			BMPs&$\tan\beta$& $\mu$ (GeV)&  $M_1$ (GeV)& $M_{2}$ (GeV)&  $M_A$ (GeV)  &$m_{\tilde{\chi}_1^0}$ (GeV)& $|\mathcal{N}_{12}|^2$  \\	
			\hline
			I& 5 & 7750 &  9130 & 2000 & 3500 & 1999 & $99.98\%$ \\
			\hline
			
			BMPs& $\Omega h^2$ &$C_{L,R}^{\rm LO}(h)$ & $C_{L,R}^{\rm NLO}(h)$& Quark \textbf{twist-2} & Gluon two-loop& $\sigma_{\rm SI}^{\rm LO}$ (pb)& $\sigma_{\rm SI}^{\rm NLO}$ (pb)\\
			\hline
			I& 0.119 & $4.5718\times 10^{-3}$ & $1.4110\times 10^{-2}$ & $-7.6326\times 10^{-10}$  &  $2.1945\times 10^{-10}$ &  $1.3710\times 10^{-11}$ & $4.1540\times 10^{-17}$  \\
			\hline
			\hline
			\multicolumn{8}{|c|}{$\mu<0$ scenario} \\ 		
			\hline
			
			BMPs&$\tan\beta$& $\mu$ (GeV)&  $M_1$ (GeV)& $M_{2}$ (GeV)&  $M_A$ (GeV)  &$m_{\tilde{\chi}_1^0}$ (GeV)& $|\mathcal{N}_{12}|^2$  \\	
			\hline
			II& 5 & $-5968$ &  4785 & 2299 & 3500 & 2299 & $99.98\%$ \\
			\hline
			
			BMPs& $\Omega h^2$ &$C_{L,R}^{\rm LO}(h)$ & $C_{L,R}^{\rm NLO}(h)$& Quark \textbf{twist-2} & Gluon two-loop& $\sigma_{\rm SI}^{\rm LO}$ (pb) & $\sigma_{\rm SI}^{\rm NLO}$ (pb)\\
			\hline
			II& 0.118 & $3.8778\times 10^{-6}$ & $9.5196\times 10^{-3}$ & $-7.6504\times 10^{-10}$  &  $2.1948\times 10^{-10}$ &  $1.8030\times 10^{-17}$ & $1.4400\times 10^{-11}$  \\
			\hline
			
		\end{tabular}
		\caption{BMPs for both $\mu>0$ and $\mu<0$ scenarios. Here, ``Quark \textbf{twist-2}" and ``Gluon two-loop" indicate the amplitudes for the quark \textbf{twist-2} contributions (Fig.~\ref{fig:boxdiag}b) and the gluon two-loop scalar contributions (Fig.~\ref{fig:twoloop}b and \ref{fig:twoloop}c), both expressed in units of ${\rm GeV}^{-2}$.
			While BMP-I registers the scenario where a \textbf{blind spot} can be realized for the NLO-improved DM-nucleon cross-section for $\mu>0$, BMP-II presents the disappearance of a LO \textbf{blind spot} with the improved cross-section for $\mu <0$. Here, all the theoretical and experimental constraints are satisfied. The DM indirect search results are not considered following the uncertainties in the density profiles.}
		\label{tab:bmplabel}
	\end{table}

	\begin{table}[h!]
		\centering
		\begin{tabular}{|c|c|c|}
			\hline
			BMPs & One-loop \big[$C_{L,R}^{\rm 1L}(h)$\big] & Counterterm \big[$\delta C_{L,R}(h)$\big] \\
			\hline
			I & $8.1201\times 10^{-3}$ & $1.4152\times 10^{-3}$ \\
			\hline
			II & $9.5857\times 10^{-3}$ & $-7.0069\times 10^{-5}$ \\
			\hline
		\end{tabular}
		\caption{One-loop and counterterm contributions to the $\tilde{\chi}_1^0\tilde{\chi}_1^0h$ vertex for the BMPs of Tab.~\ref{tab:bmplabel}.}
		\label{tab:example}
	\end{table}

The main essence of our work has been further summarized through a
few BMPs in Tab.~\ref{tab:bmplabel} and \ref{tab:example}. Here, we consider both signs for the $\mu$ parameter 
to exhibit our results.  
The BMPs satisfy the $B$-physics constraints, relic abundance data~\footnote{We have computed the relic density using the on-shell masses of the charginos and neutralinos. Alternatively, it can be evaluated using the $\overline{\rm DR}$ scheme, and consequently, the relic density value may shift.} within the $2\sigma$ variations, the Higgs mass, and other collider physics observables. 
In the $\mu>0$ scenario, BMP-I clearly asserts the fact that a minimal Higgsino part in the LSP may lead to an almost complete cancellation in the DM-nucleon NLO amplitude.  
This shows that the LO cross-section for BMP-I is reduced by a factor of $\sim \mathcal{O}(10^{-6})$ in magnitude after incorporating the NLO corrections, which results in a \textbf{blind spot} at NLO.
BMP-II presents the disappearance of the LO \textbf{blind spot} when the improved NLO cross-section is considered.

One can notice that the diagram in Fig.~\ref{fig:topology1}b offers the maximum contribution to the $\tilde{\chi}_1^0\tilde{\chi}_1^0h$ vertex at NLO. For BMP-I(BMP-II), the contribution from topology-\ref{fig:topology1}a is $\mathcal{O}(10^{-3})\left(\mathcal{O}(10^{-4})\right)$, topology-\ref{fig:topology1}b is $\mathcal{O}(10^{-2})\left(\mathcal{O}(10^{-2})\right)$, topology-\ref{fig:topology1}c is $\mathcal{O}(10^{-3})\left(\mathcal{O}(10^{-4})\right)$, topology-\ref{fig:topology1}d is $\mathcal{O}(10^{-5})\left(\mathcal{O}(10^{-5})\right)$, and the combined contribution from topology-\ref{fig:topology1}e and topology-\ref{fig:topology1}f is $\mathcal{O}(10^{-4})\left(\mathcal{O}(10^{-6})\right)$. Additionally, we have verified that for topologies \ref{fig:topology1}c and \ref{fig:topology1}d, the maximum contributions occur when squarks are present in the loop.
 
The corresponding total one-loop and the counterterm contributions are presented in Tab.~\ref{tab:example}. One can notice that the counterterm contribution in BMP-II is two orders of magnitude smaller than the corresponding one-loop contribution. On the other hand, in BMP-I, the one-loop and counterterm contributions are of the same order, as the LO $\tilde{\chi}_1^0\tilde{\chi}_1^0h$ coupling becomes quite high in this case. 
Overall, $\sigma_{\rm SI}^{\rm NLO}$ shows a profound prospect for the BMP-II, which is otherwise completely away from the present and future DD searches. 

\section{Conclusions}
\label{sec:conclusion}
We have revisited the DM-nucleon SI scattering cross-sections for the Wino-like LSP. Because the DD cross-section of Wino-like DM is exceedingly small at the LO, a systematic study of the NLO
corrections to the SI-DD cross-sections turns out to be extremely important. We begin with a brief discussion of the contributions that were considered earlier, and then, subsequently, we present the new contributions and their impacts on the SI-DD cross-section. For instance,
we have accounted for all the three-point vertices associated with the LSP-Higgs interactions involving the MSSM particles. 
Additionally, we have incorporated the available NLO EW box contributions to the $\tilde{\chi}_1^0\tilde{\chi}_1^0\bar{q}q$ process, as well as the LO and NLO contributions to the $\tilde{\chi}_1^0\tilde{\chi}_1^0gg$ processes in $\mathtt {MicrOMEGAs}$. Through a comprehensive numerical scan across the parameter space, we have explored scenarios in which DM may exist primarily as a Wino-like state with as large as $10\%$ Higgsino component. 
The total cross-sections receive a significant cancellation between the quark \textbf{twist-2} and scalar contributions involving quarks and gluons, making it smaller than the LO values in most parts of the parameter space for $\mu>0$. Furthermore, when the LSP contains a small but non-zero Higgsino fraction of $\sim\mathcal{O}(0.02\%)$, the NLO cross-sections decrease significantly compared to their LO values, resulting in the appearance of \textbf{blind spots}. 
In contrast, in the $\mu<0$ scenario, a relative sign difference between the Wino and Higgsino mass terms can cause the LO Wino-Higgsino-Higgs coupling to vanish, resulting in extremely small LO cross-sections. This produces \textbf{blind spots} at LO. 
However, in this scenario, NLO corrections can enhance the cross-sections, thereby effectively eliminating the \textbf{blind spots} at NLO.
Therefore, the loop-corrected cross-sections have been significantly shifted upwards to reach the DD bounds from the \textbf{LZ} experiment, making it potentially detectable in the next generation of DD experiments for DM searches. 

\section{ACKNOWLEDGEMENTS}
The computations were supported in part by SAMKHYA, the High-Performance Computing Facility
provided by the IoP, Bhubaneswar, India. The authors acknowledge useful discussion with A. Pukhov. S.B. acknowledges N. Nagata for the valuable discussions. AC acknowledges the hospitality at IoP, Bhubaneswar, during the meeting IMHEP-19 and IMHEP-22, which facilitated this work. AC and SAP also acknowledge the hospitality at IoP, Bhubaneswar, during a visit. SB acknowledges the local hospitality at SNIoE, Greater Noida, during the meeting at WPAC-2023, where this work was put in motion. S.B. and D.D. also acknowledge the local hospitality received at SNIoE, during the final stage of the work.

\bigskip
\bibliographystyle{JHEPCust.bst}
\bibliography{Wino_DM}
\end{document}